\documentclass[ 
    aps,
    pra,
    twocolumn,
    letterpaper, 
    10pt,
    superscriptaddress, 
    showpacs,
    showkeys,
    notitlepage,
    amsmath, 
    amssymb, 
    floatfix 
]{revtex4-1} 
 
\usepackage{graphicx}% Include figure files
\usepackage{dcolumn}% Align table columns on decimal point
\usepackage{bm}% bold math
\usepackage{color}
\usepackage{amssymb}
\usepackage{amsmath}
\usepackage{hyperref}% add hypertext capabilities
\usepackage{bbold}
\usepackage{accents}

\newcommand{\harpvecsign}{\scriptscriptstyle\text{\tiny$\leftrightarrow$}}
\newcommand{\harpoonvec}[2]{%
  \ifx\displaystyle#1\doalign{$\harpvecsign$}{#1#2}\fi
  \ifx\textstyle#1\doalign{$\harpvecsign$}{#1#2}\fi
  \ifx\scriptstyle#1\doalign{\scalebox{.6}[.9]{$\harpvecsign$}}{#1#2}\fi
  \ifx\scriptscriptstyle#1\doalign{\scalebox{.5}[.8]{$\harpvecsign$}}{#1#2}\fi
}
\newcommand{\doalign}[2]{%
 {\vbox{\offinterlineskip\ialign{\hfil##\hfil\cr#1\cr$#2$\cr}}}%
}

\newcommand{\bel}{\begin{equation}}
\newcommand{\eel}{\end{equation}}

\newcommand{\skyp}[1]{}

\newcommand{\fr}{\frac}

\newcommand{\vare}{\varepsilon}

\newcommand{\ee}{\end{equation}}
\newcommand{\be}{\begin{equation}}
\newcommand{\mbf}{\mathbf}

\newcommand{\bal}{\begin{eqnarray} }
\newcommand{\eal}{\end{eqnarray}}
\newcommand{\ba}{\begin{eqnarray*}}
\newcommand{\ea}{\end{eqnarray*}}

\newcommand{\reffig}[1]{Fig.~\ref{#1}}

\newcommand{\ket}[1]{| #1 \rangle}
\newcommand{\bra}[1]{\langle #1 |}

\newcommand{\bq}{{\mathbf q}}
\newcommand{\bG}{{\mathbf G}}
\newcommand{\bp}{{\mathbf p}}
\newcommand{\brho}{{\boldsymbol \rho}}
\newcommand{\br}{{\mathbf r}}

\newcommand{\bR}{{\mathbf R}}

\newcommand{\bk}{{\mathbf k}}
\newcommand{\bK}{{\mathbf K}}

\newcommand{\pa}{\partial}

\newcommand{\refeq}[1]{Eq.~\eqref{#1}}

\begin{document} 

%%%%%%%%%%%%%%%%%%%%%%%%%%%%%%%%%%%%%%%%
%%%%%%%%%%%%%%%%%%%%%%%%%%%%%%%%%%%%%%%%
\title{ Photonic Band Structure of Two-dimensional Atomic Lattices}

\author{J. Perczel}
\affiliation{Physics Department, Massachusetts Institute of Technology, Cambridge, MA 02139, USA}
\affiliation{Physics Department, Harvard University, Cambridge,
MA 02138, USA}

\author{J. Borregaard}
\affiliation{Physics Department, Harvard University, Cambridge,
MA 02138, USA}
\affiliation{QMATH, Department of Mathematical Sciences, University of Copenhagen,  2100 Copenhagen \O, Denmark}

\author{D. E. Chang}
\affiliation{ICFO - Institut de Ciencies Fotoniques, The Barcelona Institute of Science and Technology, 08860 Castelldefels, Barcelona, Spain}

\author{H. Pichler}
\affiliation{Physics Department, Harvard University, Cambridge,
MA 02138, USA}
\affiliation{ITAMP, Harvard-Smithsonian Center for Astrophysics, Cambridge, MA 02138, USA}

\author{S. F. Yelin}
\affiliation{Physics Department, Harvard University, Cambridge,
MA 02138, USA}
\affiliation{Department of Physics, University of Connecticut, Storrs, Connecticut 06269, USA}

\author{P. Zoller}
\affiliation{Institute for Theoretical Physics, University of Innsbruck, A-6020 Innsbruck, Austria}
\affiliation{Institute for Quantum Optics and Quantum Information of the Austrian Academy of Sciences, A-6020 Innsbruck, Austria}

\author{M. D. Lukin}
\affiliation{Physics Department, Harvard University, Cambridge,
MA 02138, USA}

% \author{Authors}

\date{\today}

\bigskip
\bigskip
\bigskip
%%%%%%%%%%%%%%%%%%%%%%%%%%%%%%%%%%%%%%%%%%%%%%%%%%%%%%%%%%%%%%%%%%%%%%%%%%%%%%%%%%5
%%%%%%%%%%%%%%%%%%%%%%%%%%%%%%%%%%%%%%%%%%%%%%%%%%%%%%%%%%%%%%%%%%%%%%%%%%%%%%%%%%5
% Abstract
%%%%%%%%%%%%%%%%%%%%%%%%%%%%%%%%%%%%%%%%%%%%%%%%%%%%%%%%%%%%%%%%%%%%%%%%%%%%%%%%%%5
%%%%%%%%%%%%%%%%%%%%%%%%%%%%%%%%%%%%%%%%%%%%%%%%%%%%%%%%%%%%%%%%%%%%%%%%%%%%%%%%%%5
\begin{abstract} 
%Recently, two-dimensional atomic arrays have been shown to exhibit intriguing topological phenomena at optical frequencies that persist even when atomic coupling to free-space modes is fully accounted for. 

Two-dimensional atomic arrays exhibit a number of intriguing quantum optical phenomena, including subradiance, nearly perfect reflection of radiation and 
long-lived topological edge states. Studies of emission and scattering of photons in such lattices require complete treatment of the radiation pattern from individual atoms, including long-range interactions. 
%Accurately calculating  is a challenging task. 
We describe a systematic approach to perform the calculations 
of collective energy shifts and decay rates in the presence of such long-range interactions for arbitrary two-dimensional atomic lattices. As applications of our method, we investigate the topological properties of atomic lattices both in free-space and near plasmonic surfaces.

\end{abstract}

\maketitle
 
%%%%%%%%%%%%%%%%%%%%%%%%%%%%%%%%%%%%%%%%
%%%%%%%%%%%%%%%%%%%%%%%%%%%%%%%%%%%%%%%%

%%%%%%%%%%%%%%%%%%%%%%%%%%%%%%%%%%%%%%%%%%%%%%%%%%%%%%%%%%%%%%%%%%%%%%%%%%%%%%%%%%5
%%%%%%%%%%%%%%%%%%%%%%%%%%%%%%%%%%%%%%%%%%%%%%%%%%%%%%%%%%%%%%%%%%%%%%%%%%%%%%%%%%5
% Introduction 
%%%%%%%%%%%%%%%%%%%%%%%%%%%%%%%%%%%%%%%%%%%%%%%%%%%%%%%%%%%%%%%%%%%%%%%%%%%%%%%%%%5
%%%%%%%%%%%%%%%%%%%%%%%%%%%%%%%%%%%%%%%%%%%%%%%%%%%%%%%%%%%%%%%%%%%%%%%%%%%%%%%%%%5

%%%%%%%%%%%%%%%%%%%%%%%%%%%%%%%%
\section{Introduction}
%%%%%%%%%%%%%%%%%%%%%%%%%%%%%%%%

Quantum optical properties of lattices of atoms and atom-like emitters are being actively explored both theoretically and experimentally \cite{Deutsch1995,VanCoevorden1996,DeVries1998,Klugkist2006,Antezza2009,Antezza2009a,Bienaime2012,Guerin2016,Perczel2017,Bettles2017,Bettles2016,Syzranov2016,Shahmoon2017,Asenjo-Garcia2017}. In such lattices, atoms are assumed to be confined such that tunneling between sites is negligible and they interact via photon-mediated dipole-dipole interactions giving rise to hybridized atom-photon bands. 
The photonic band structure of three-dimensional (3D) atomic lattices has been 
investigated in a number of studies \cite{VanCoevorden1996,DeVries1998,Klugkist2006,Antezza2009,Antezza2009a}. Recently, there has been significant interest in the photonic properties of two-dimensional (2D) atomic lattices, which have been shown to exhibit a variety of remarkable phenomena, including subradiance \cite{Facchinetti2016,Asenjo-Garcia2017}, near perfect reflection of radiation \cite{Bettles2016,Shahmoon2017} and long-lived topological excitations \cite{Perczel2017,Bettles2017}.

In such lattices, a key distinction arises between the radiative interatomic coupling that gives rise to collective behavior and the radiative coupling to free-space modes that leads to unwanted losses.
 %and efficient photon storage \cite{Asenjo-Garcia2017}. 
%Such 2D lattices have recently been recognized as a promising platform. Recently, it was shown that two-dimensional periodic atomic arrays give rise to remarkable topological properties at optical frequencies, including photonic bands with non-zero Chern numbers, long-lived topological bound-states and unidirectional emission of individual atoms near the system boundary \cite{Perczel2017}. These phenomena survive in subwavelength lattices even when coupling of atoms to free space modes is fully accounted for. In these lattices atoms are assumed to be tightly trapped at the lattice sites such that tunneling between sites is negligible and the atoms interact via dipole-dipole interactions giving rise to hybridized atom-photon bands. %The photonic band structure of such atomic lattices has been investigated in a number of works over recent years \cite{Klugkist2006,Antezza2009,Antezza2009a,Bettles2016,Shahmoon2017}.  
%Recently, such 2D systems have also been shown to form atomically thin mirrors \cite{Bettles2016,Shahmoon2017} and to be capable of storing photons \cite{Asenjo-Garcia2017}. 
In order to fully account for the radiative loss of atoms, the atomic coupling to all free-space modes, including propagating long-range photons, has to be considered. Determining the eigenmodes of the lattice in the presence of such long-range interactions between atoms is a non-trivial task, requiring careful treatment in order to obtain accurate results and to understand the photonic properties of the lattice.

In this work, we describe a general approach for the calculation of photonic band structures in two-dimensional atomic lattices with arbitrary lattice geometries. %We consider atomic transitions with all three polarizations. 
Previously, general methods have been put forward ~\cite{VanCoevorden1996,DeVries1998,Klugkist2006,Antezza2009,Antezza2009a} to accurately calculate photonic band structures for 3D infinite atomic lattices. %, where the system is periodic in all directions. 
Here, we make use of the general approach 
described in Refs.~\cite{Antezza2009,Antezza2009a} and 
focus on 
%specialize the methods  to describe 
2D lattices, where the axial symmetry is broken along the third dimension. Previous  calculations involving infinite 2D lattices were either restricted to two-level atoms in a square lattice \cite{Asenjo-Garcia2017}, required summations in real-space where convergence is slow \cite{Shahmoon2017,Bettles2017}, or relied on the method that we will now describe in detail \cite{Perczel2017}. As an application of our method, we study the topological properties of both Bravais and non-Bravais lattices in free space. 

 %%%%%%%
\begin{figure}[h!]
\centering
\includegraphics[width=0.25 \textwidth]{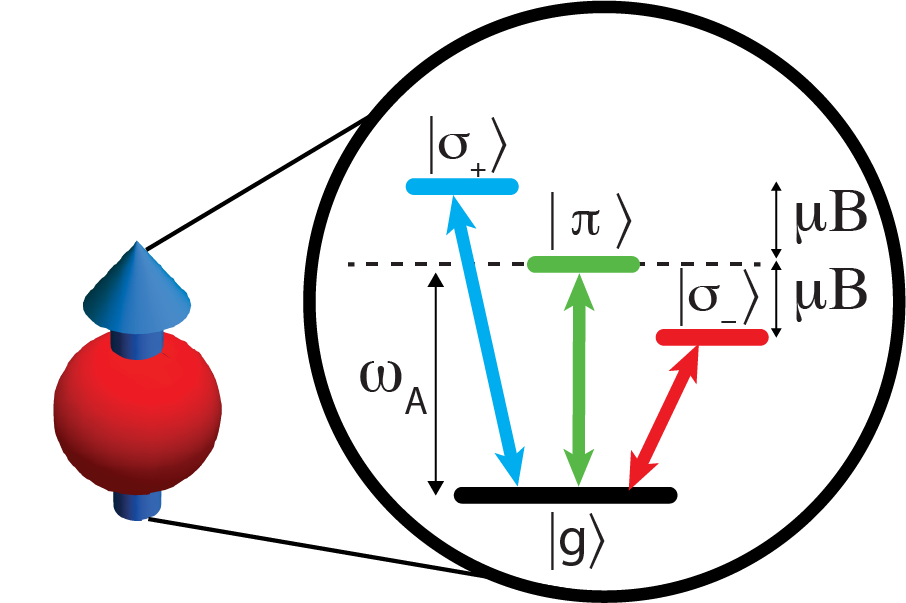}
\caption{
\label{atom_internal}
Constituent atoms in  the 2D atomic lattice. Each atom is assumed to have one ground $\ket{g}$ and three excited states $\ket{\sigma_+}$, $\ket{\sigma_-}$ and $\ket{\pi}$ that can be Zeeman split by a magnetic field. }
\end{figure} 
%%%%%%%

Furthermore, our formalism can also be used to describe two-dimensional lattices of emitters near planar surfaces. As a novel application, we study atomic lattices close to plasmonic surfaces and describe their emerging topological character. We note that studying the emission pattern of dipoles near metallic surfaces has been an active research area for over a century \cite{Collin2004} and has been at the forefront of plasmonics research over the past few years \cite{Ebbesen1998,Chang2006,GarciadeAbajo2007,Nikitin2010,Kildishev2013,Tame2013,Delga2014,High2015}. Our work provides a new and simple framework to study the collective decay rates into both free-space modes and plasmonic channels of a periodic lattice of dipole-like emitters near a metal surface, while also taking into account realistic metallic dispersion.

Our manuscript  is organized as follows. In section II we describe a general analytical approach to finding the Bloch modes of an infinite two-dimensional atomic lattice. In section III we apply our formalism to analyze a non-Bravais square lattice and a triangular lattice of atoms in free space and discuss their topological properties. In section IV we discuss a square lattice and a triangular lattice of atoms in the vicinity of a silver surface and discuss the topological properties of the latter. Key results and conclusions are summarized in section V.

%%%%%%%%%%%%%%%%%%%%%%%%%%%%%%%%
\section{General formalism}
%%%%%%%%%%%%%%%%%%%%%%%%%%%%%%%%

We consider a 2D lattice of atoms in the $x$-$ y$ plane with interatomic spacing $a$. The quantization axis $\hat z$ is perpendicular to the plane of the atoms. Each atom is assumed, for simplicity, to have transitions from the ground state $\ket{g}$ to the excited states ${\ket{\sigma_\pm}=\mp(\ket{x}\pm i\ket{y})/\sqrt{2}}$ and $\ket{\pi}=\ket{z}$, which are excited by $\hat \sigma_\pm$ and $\hat z$ polarized light respectively (Fig.~\ref{atom_internal}). Note that our formalism can be extended in a straightforward way to describe atoms with different internal level structures as long as only a single ground state is considered.

%%%%%%%%%%%%%%%%%%%%%%%%%%%%%%%%
\subsection{Hamiltonian}\label{HamiltonianSection}
%%%%%%%%%%%%%%%%%%%%%%%%%%%%%%%%
We use the dipole approximation to write down the Hamiltonian describing the interaction of the atoms with the quantized radiation field \cite{Antezza2009,Antezza2009a,Pichler2015}
\bal\label{Hamiltonian}
&&H = \hbar \sum\limits_{n=1}^N  \omega_A\Big (\ket{\sigma_{+,n}}\bra{\sigma_{+,n}}+\ket{\sigma_{-,n}}\bra{\sigma_{-,n}}+\ket{\pi_n}\bra{\pi_n}\Big) \nonumber \\
&&\;+\int d^3k\sum\limits_{\epsilon}\hbar c k a_{\bk \epsilon}^\dagger  a_{\bk \epsilon} - \sum\limits_{n=1}^{N}\mbf{d}_n\cdot \mbf{E}(\br_n)+H_\text{Zeeman},
\eal
where $N$ is the number of atoms, $\omega_A=2\pi c/\lambda$ is the atomic transition frequency with wavelength $\lambda$ and $c$ is the speed of light in vacuum. Here $\ket{\sigma_{\pm,n}}\bra{\sigma_{\pm,n}}$ and $\ket{\pi_n}\bra{\pi_n}$ represent operators that only act on the subspace of the $n$-th atom. The creation and annihilation operators of the electromagnetic field satisfy ${[ a_{\bf k\epsilon},a_{\bk',\epsilon'}^\dagger]=\delta_{\epsilon,\epsilon '}\delta(\bk - \bk')}$, where $\bk$ is the wave vector with magnitude $k=|\bk|$ and $\epsilon$ labels the two photon polarizations $\hat \epsilon_{\bk}$ perpendicular to $\bk$. The atomic transition dipole operator is given by ${\mbf d = d(\ket{\sigma_+}\bra{g}\hat \sigma_++\ket{\sigma_-}\bra{g}\hat \sigma_-+\ket{\pi}\bra{g}\hat z)+\text{h.c.}}$, where we assume, for simplicity, that all three transition dipole moments are equal. The dipole operator couples to the quantized transverse electromagnetic field modes ${\mbf E(\br) = \int d^3k\sum_{\epsilon}[\mathcal E_k\hat \epsilon_\bk  a_{\bk \epsilon} e^{i\bk\cdot \br}+\text{h.c.}]}$, where ${\mathcal E_k = i(2\pi)^{-3/2}[\hbar k c/(2\epsilon_0)]^{1/2}}$ and $\br_n$ denotes the position vector of individual atoms. The Hamiltonian accounting for the Zeeman splitting of the atoms is given by
\bal
H_\text{Zeeman} = \hbar\sum\limits_{n=1}^N\mu B\big( \ket{\sigma_{+,n}}\bra{\sigma_{+,n}}-\ket{\sigma_{-,n}}\bra{\sigma_{-,n}}\big),
\eal
where $\mu B$ is the Zeeman shift of the atomic levels with magnetic moment $\mu$ due to an out-of-plane magnetic field ${\mbf B = B\hat z}$.

Following the adiabatic elimination of the reservoir degrees of freedom via the Born-Markov approximation \cite{VanCoevorden1996,DeVries1998,Klugkist2006,Antezza2009,Antezza2009a,Gardiner2010,Pichler2015,Asenjo-Garcia2017}, we obtain a master equation for the evolution of the system density operator $\rho(t)$, which in the single excitation sector is given by
%\begin{widetext}
\bal\label{master}
&&\dot \rho = - \fr{i}{\hbar}\left( H_\text{eff}\rho - \rho H_\text{eff}^\dagger \right)\\
&&+\fr{6\pi \hbar\Gamma_0c}{\omega_A}\sum\limits_{i,j=1}^{N}\sum\limits_{\alpha,\beta=\sigma_+,\sigma_-,\pi}\kern-1em \!\text{Im} \;G_{\alpha\beta}(\br_i-\br_j)\ket{g_i}\bra{\alpha_i}\rho\ket{\beta_j}\bra{g_j},\nonumber
\eal
%\end{widetext}
where the non-Hermitian, effective spin Hamiltonian reads
%\begin{widetext}
\bal\label{Hamiltonian}
H_\text{eff}&=&\hbar\sum\limits_{i=1}^{N}\sum\limits_{\alpha=\sigma_+,
\sigma_-,\pi}\left(\omega_A-\text{i}\fr{\Gamma_0}{2}\right)\ket{\alpha_i}\bra{\alpha_i}+H_\text{Zeeman}\nonumber \\
&+&\fr{3\pi \hbar\Gamma_0c}{\omega_A}\sum\limits_{i\neq j}\sum\limits_{\alpha,\beta=\sigma_+,
\sigma_-,\pi}\kern-1em G_{\alpha\beta}(\br_i-\br_j)\ket{\alpha_i}\bra{\beta_j},\quad
\eal
%\bal\label{Hamiltonian}
%H_\text{eff}=\hbar\sum\limits_{i=1}^{N}\sum\limits_{\alpha=x,y,z}\left(\omega_A-\text{i}\fr{\Gamma_0}{2}\right)\ket{\alpha_i}\bra{\alpha_i}+\fr{3\pi \hbar\Gamma_0c}{\omega_A}\sum\limits_{i\neq j}\sum\limits_{\alpha,\beta=x,y,x}G_{\alpha\beta}(\br_i-\br_j)\ket{\alpha_i}\bra{\beta_j},\quad
%\eal
%\end{widetext}
and $\Gamma_0=d^2\omega_A^3/(3\pi\epsilon_0\hbar c^3)$ is the radiative linewidth of a single atom in free space and $G_{\alpha\beta}(\br)$ is the dyadic Green's function describing the dipolar spin-spin interactions (see Section \ref{greens}). Note that as part of the Markov approximation, the only frequency-dependence is through the atomic frequency $\omega_A$ \cite{Asenjo-Garcia2017}.

The first term on the right-hand side of \refeq{master} describes the deterministic evolution of the atomic wavefunction and includes dipole-dipole interactions mediated via photon exchange, whereas the second term accounts for stochastic quantum jumps \cite{Dalibard1992,Carmichael1993, Molmer1993,Asenjo-Garcia2017}. In the absence of a driving field, the dynamics in the single excitation sector is completely characterized by the non-Hermitian Hamiltonian in \refeq{Hamiltonian}, since a quantum jump prepares the system in a trivial state where all atoms are in their ground states and the system does not evolve. Therefore, it is not necessary to keep track of these quantum jumps and the system dynamics can be studied without including the second term on the right-hand side of \refeq{master}. The time evolution of the system is then described by the non-Hermitian Hamiltonian in \refeq{Hamiltonian} via the equation
\be\label{schroedinger}
H_\text{eff}\ket{\psi(t)}=i\hbar\partial_t\ket{\psi(t)},
\ee
where the overall decrease in amplitude reflects the dissipative transfer of population to the ground state.

%Before proceeding, we note that in our system the atom-atom interactions arise from frequency-conserving, elastic scattering of photons from the atoms. These interactions depend on the frequency of the photons that mediate the dipole-dipole interaction [${J_{ij}(\omega)\sim \omega^2G_{\alpha\beta}(\omega,\br_i-\br_j)}$]. Therefore, in general, the effective Hamiltonian itself depends on frequency and the following implicit matrix equation $H_\text{eff}(\omega)\ket{\psi}=\hbar \omega \ket{\psi}$ has to be solved to find the eigenmodes of the system while enforcing self-consistency. However, in the Markov regime both $\omega$ and $\omega_A$ are much larger than $\Gamma_0$, which is the only relevant energy scale of our problem. Thus we may replace $\omega$ with $\omega_A$ everywhere in the Hamiltonian to obtain Eq.~\eqref{Hamiltonian} \cite{Asenjo-Garcia2017}. 

%%%%%%%%%%%%%%%%%%%%%%%%%%%%%%%%
\subsection{Bravais lattices}
%%%%%%%%%%%%%%%%%%%%%%%%%%%%%%%%

For an infinite periodic Bravais lattice, which has a single atom per unit cell, the single excitation eigenmodes of \refeq{Hamiltonian} are Bloch modes of the form
\bal\label{BravaisBlochmodes}
\ket{\psi_{\bk_B}}=\sum\limits_n e^{i\bk_B\cdot \bR_n}\Big(c_{+}\ket{\sigma_{+,n}}+c_{-}\ket{\sigma_{-,n}}+c_{z}\ket{z_n}\Big),\qquad
\eal
where the summation runs over all lattice vectors $\{\bR_n\}$, $\bk_B$ is the Bloch wavevector, and $c_{+}$, $c_{-}$ and $c_z$ are constants that depend, in general, on ${\bk_B}$. 

It is convenient to solve for and manipulate the dyadic Green's function in the Cartesian basis. Therefore, we transform \refeq{Hamiltonian} and \refeq{BravaisBlochmodes} using the relation ${\ket{\sigma_\pm}=\mp(\ket{x}\pm i\ket{y})/\sqrt{2}}$ and perform all calculations in the Cartesian basis in the rest of this paper. 

Using the Bloch ansatz for the wavefunction, finding the eigenvalues $E_{\textbf{k}_B}$ of the effective Hamiltonian $H_{\rm eff}$ reduces to diagonalizing the following $3\times3$ matrix $\mbf M$, whose components are given by 
\be\label{energiesBravais}
\mbf M_{\alpha\beta} =\left(\omega_A-i\Gamma_0/2\right)\delta_{\alpha\beta}+\xi_{\alpha\beta}+\chi_{\alpha\beta},
\ee
where $\alpha,\beta = x,y,z$ label the polarization components, and $\delta_{\alpha\beta}$ is the Kronecker delta. Here $\xi_{\alpha\beta}$ and $\chi_{\alpha\beta}$ stand for the components of the complex matrices accounting for the magnetic field and the atom-atom interactions, respectively, and take the form
\bal
\xi_{\alpha\beta} = - i\mu B(\delta_{\alpha x}\delta_{\beta y}-\delta_{\alpha y}\delta_{\beta x}),
\eal
and
\be\label{chiBravais}
\chi_{\alpha\beta} = \fr{3\pi \Gamma_0c}{\omega_A} \sum\limits_{\bR\neq 0} e^{i\bk_B \cdot \bR}G_{\alpha\beta}(\bR).
\ee
The matrix $\mbf M$ captures how the energy levels, decay rates, and internal level couplings of a single atom are affected by the magnetic field and the presence of all the other atoms in the periodic lattice for a given Bloch vector $\bk_B$. For example, the matrix element $\chi_{xy}$ captures how the $\ket{x}$ state of an individual atom is affected by the couplings to the $\ket{y}$ states of all the other atoms in the lattice. The diagonalization of $\mbf M$ yields three complex eigenvalues for each value of the Bloch vector $\bk_B$ of the form $E_{\bk_B}=\omega_{\bk_B} - i\gamma_{\bk_B}$, where the real part $\omega_{\bk_B}$ corresponds to the energy of the Bloch eigenmode and the imaginary part $\gamma_{\bk_B}$ gives the overall decay rate of the mode. %Additionally, we avoid the frequency-dependence of the Hamiltonian by replacing $\omega$ by $\omega_A$ (see discussion at the end of Section \ref{HamiltonianSection}).

%%%%%%%%%%%%%%%%%%%%%%%%%%%%%%%%
\subsection{Non-Bravais lattices}
%%%%%%%%%%%%%%%%%%%%%%%%%%%%%%%%

For an infinite periodic non-Bravais lattice, with $m$ sites per unit cell, the single excitation eigenmodes of \refeq{Hamiltonian} are Bloch modes of the form \cite{Bena2009}
\bal\label{NBwavefunction}
\ket{\psi}=\sum\limits_n \sum\limits_{b=1}^me^{i\bk_B\cdot \bR_n}\Big(c_{+}^{b}\ket{\sigma_{+,n}^{b}}+c_{-}^{b}\ket{\sigma_{-,n}^{b}}+c_{z}^{b}\ket{z_{n}^{b}}\Big)\!,\quad\;\;\;
\eal
where $b$ labels the different atoms within the unit cell. In this case, the eigenmodes of the system are obtained by diagonalizing a $3m\times 3m$ matrix.

For simplicity, we focus on non-Bravais lattices with two sites per cell, but the formalism can be extended in a straightforward way to include more sites per cell. For $m=2$, the non-Bravais lattice can be thought of as a lattice arising from the union of two sublattices ${\{R_1\}}$ and ${\{R_2\}}$, which are shifted with respect to each other by the basis vector $\mbf b$ that points from one site to the other within the periodic unit cell. With this notation, the matrix components of the relevant $6\times 6$ complex matrix $\mbf M$ are given by
\bal\label{energiesNonBravais}
\mbf M_{\alpha\mu,\beta\nu} &=&\left(\omega_A^{(1)}-i\Gamma_0/2\right)\delta_{\alpha\beta}\delta_{1\mu}\delta_{1\nu}+ \xi_{\alpha\mu,\beta\nu}\nonumber\\
&+&\left(\omega_A^{(2)}-i\Gamma_0/2\right)\delta_{\alpha\beta}\delta_{2\mu}\delta_{2\nu}+\chi_{\alpha\mu,\beta\nu},
\eal
where $\omega_A^{(1)}$ and $\omega_A^{(2)}$ are the transition frequencies of the atoms located on the two sublattices, while $\mu$ and $\nu$ are sublattice labels than run over $\mu,\nu=1,2$. The Terms accounting for the magnetic field interaction are given by
\bal
\xi_{\alpha\mu,\beta\nu} = - i\mu B(\delta_{\alpha x}\delta_{\beta y}-\delta_{\alpha y}\delta_{\beta x})(\delta_{1\mu}\delta_{1\nu}+ \delta_{2\mu}\delta_{2\nu}),\qquad
\eal
and the terms describing the atom-atom interactions take the form
\bal\label{chiNonBravais}
\chi_{\alpha\mu,\beta\nu} = \fr{3\pi \Gamma_0c}{\omega_A^{(1)}} &\Bigg[&\sum\limits_{\bR_1\neq 0} e^{i\bk_B \cdot \bR_1}G_{\alpha\beta}(\bR_1)\delta_{1\mu}\delta_{1\nu}\quad\nonumber\\
&+&\sum\limits_{\bR_1} e^{i\bk_B \cdot \bR_1}G_{\alpha\beta}(\bR_1+\mbf b)\delta_{1\mu}\delta_{2\nu}\nonumber\\
&+&\sum\limits_{\bR_2\neq 0} e^{i\bk_B \cdot \bR_2}G_{\alpha\beta}(\bR_2)\delta_{2\mu}\delta_{2\nu}\nonumber\\
&+&\sum\limits_{\bR_2} e^{i\bk_B \cdot \bR_2}G_{\alpha\beta}(\bR_2-\mbf b)\delta_{2\mu}\delta_{1\nu}\Bigg],\quad\;\;\;\;
\eal
where we have used the fact that $|\omega_A^{(1)}-\omega_A^{(2)}|/\omega_A^{(1)}\ll 1$ to replace all occurrences of $\omega_A^{(2)}$ with $\omega_A^{(1)}$ in $\chi_{\alpha\mu,\beta\nu}$ and factor out a common prefactor.
%\footnote{We assume, for simplicity, that ${\Gamma_0\sim [\omega_A^{(1)}]^3}$ for all atoms. For optical frequencies (${\Gamma_0\ll \omega_A^{(1)},\omega_A^{(2)}}$) the difference between ${\Gamma_0(\omega_A^{(1)})}$ and ${\Gamma_0(\omega_A^{(2)})}$ is negligible.}.
%The polarization indices run over $\alpha,\beta=x,y,z$ and the sublattice indices run over $\mu,\nu=1,2$. 
In \refeq{chiNonBravais} the first term on the right-hand side describes how atoms in sublattice ${\{R_1\}}$ affect each other, whereas the second term describes how atoms in sublattice ${\{R_1\}}$ are affected by atoms in sublattice ${\{R_2\}}$. The third and fourth terms can be interpreted similarly.

\bigskip

In principle, the eigenmodes of the lattice can be directly calculated from Eqs.~\eqref{energiesBravais} and \eqref{chiBravais} for Bravais lattices, and Eqs.~\eqref{energiesNonBravais} and \eqref{chiNonBravais} for non-Bravais lattices for arbitrary lattice geometries using the expression for the Green's function in real space and summing over all lattice sites. However, in the presence of long range interactions, as for example in free space, the summation over the lattice sites converges very slowly, making accurate numerical computations difficult. Furthermore, in certain geometries, e.g. near planar surfaces (see Section \ref{planarGreens}), no closed form expression exists for the Green's function in real space. Below we describe a method to perform the relevant summations %present in \refeq{chiBravais} and \refeq{chiNonBravais} 
in momentum space, where convergence is fast and the expression for the Green's function in momentum-space can be used for the calculation, which is typically easier to obtain than the equivalent expression in real space. 

%%%%%%%%%%%%%%%%%%%%%%%%%%%%%%%%
\subsection{Dyadic Green's function}\label{greens}
%%%%%%%%%%%%%%%%%%%%%%%%%%%%%%%%

The dyadic Green's function $G_{\alpha\beta}$ that appears in Eqs.~\eqref{master}, \eqref{Hamiltonian}, \eqref{chiBravais} and \eqref{chiNonBravais} is the solution of the dyadic equation \cite{Dung1998}
\bal\label{greensEquation}
\vare(\omega,\br)\fr{\omega^2}{c^2} G_{\alpha\beta}(\br,\br')\qquad\qquad\qquad\qquad\qquad\qquad\qquad\nonumber\\
-\left(\pa_\alpha\pa_\nu-\delta_{\alpha\nu}\pa_\eta\pa_\eta\right)G_{\nu\beta}(\br,\br')=\delta_{\alpha\beta}\delta(\br-\br'),\qquad
\eal
where $\br=x\hat x+y\hat y+z\hat z$ and $r=|\br|$. Summation is implied over repeated indices. The dielectric permittivity $\vare(\omega,\br)$ is potentially spatially inhomogeneous and frequency-dependent and we assume a non-magnetic medium with magnetic permeability ${\mu(\omega,\br)=1}$. Physically, the Green's function describes the electromagnetic radiation at position $\br$ emitted by a point-like dipole oscillating at frequency $\omega$ at position $\br'$. 

In free space, the permittivity is $\vare = 1$ and \refeq{greensEquation} has a closed-form solution for the free-space dyadic Green's function $G_{0,\alpha\beta}$ (see Appendix \ref{GreensAppendix}). The components of the Green's function with radiating boundary conditions are given by \cite{Dung1998,Morice1995}
\bal\label{freespaceGreensRealSpace}
&&G_{0,\alpha\beta}(\br)=-\fr{e^{ikr}}{4\pi r}\bigg[\bigg( 1+\fr{i}{kr}-\fr{1}{(kr)^2} \bigg)\delta_{\alpha\beta}\qquad\nonumber\\
&&\qquad+\bigg(-1-\fr{3i}{kr}+\fr{3}{(kr)^2}\bigg) \fr{x_\alpha x_\beta}{r^2}  \bigg]+\fr{\delta_{\alpha\beta}\delta^{(3)}(\br)}{3k^2},\qquad
\eal
where $k=\omega/c$ and we have used the fact that the Green's function only depends on $\br-\br'$ to write it with a single argument. This is the well-known expression describing dipole-dipole interactions in free space, which can also be derived using conventional quantum optical techniques \cite{Gross1982,Milonni1974}.

In the presence of planar interfaces, the expression for the dyadic Green's function gets more complicated and the components can be evaluated in a closed form only in momentum space. The explicit expressions for the momentum-space components of the Green's function near planar surfaces are described in detail in Appendix \ref{greensNearFlatSurface}.

%%%%%%%%%%%%%%%%%%%%%%%%%%%%%%%%
\subsection{Summation in momentum space}\label{Poisson}
%%%%%%%%%%%%%%%%%%%%%%%%%%%%%%%%

From \refeq{freespaceGreensRealSpace} it is clear that the Green's function decays as $\sim 1/r$ in the far-field limit. In the presence of such long-range interactions between atoms %, the summations in Eqs.~\eqref{chiBravais} and \eqref{chiNonBravais} over all lattice vectors converge slowly, making accurate numerical calculations challenging. Instead, 
it is desirable to perform the summations in \refeq{chiBravais} and \refeq{chiNonBravais} in momentum space, where all sums converge rapidly as previously noted.  The summation in position space is transformed to a summation in momentum space using the following form of Poisson's identity
\be\label{PoissonTransform}
\sum\limits_{\bR}e^{i(\bp+\bk_B)\cdot \bR}=\fr{1}{\mathcal A}\sum\limits_{\mbf G}(2\pi)^2\delta^{(2)}(\bp+\bk_B-\mbf G),
\ee
where  $\mathcal A$ is the area of the unit cell and the reciprocal lattice vectors $\{\mbf G\}$ are related to the lattice vectors $\{\mbf R\}$ by ${\mbf G\cdot \mbf R = 2\pi m}$ for integer $m$ \cite{Ashcroft1976}. Making use of \refeq{PoissonTransform}, we obtain 
\bal\label{transformation}
&& \sum\limits_{\bR\neq 0} e^{i\bk_B \cdot \bR}G_{\alpha\beta}(\bR)= \sum\limits_{\bR} e^{i\bk_B \cdot \bR}G_{\alpha\beta}(\bR)-G_{\alpha\beta}(\mbf 0)\nonumber\\
 &&\qquad\qquad= \fr{1}{\mathcal A}\sum\limits_{\mbf G}g_{\alpha\beta}(\mbf G-\bk_B; 0)-G_{\alpha\beta}(\mbf 0),
\eal
where we have used the \textit{Weyl decomposition} of the Green's function in terms of 2D plane waves \cite{Chew1995}, which is defined via
\be\label{Weyl}
G_{\alpha\beta}(\boldsymbol \br)=\int \fr{dp_x dp_y}{(2\pi)^2}g_{\alpha\beta}(\bp; z)e^{i\bp\cdot\mbf \brho},
\ee 
where $\bp =p_x\hat x+p_y\hat y+p_z\hat z$ and $\brho = \rho_x\hat x + \rho_y \hat y$ denotes the polar radial coordinate in the $x$-$y$ plane. In free space the Weyl decomposition is given by (see Appendix \ref{GreensAppendix})
\bal\label{pz}
g_{\alpha\beta}(\bp; z)=\int \fr{dp_z}{2\pi}e^{ip_zz}\fr{1}{k^2}\fr{k^2\delta_{\alpha\beta}-p_\alpha p_\beta}{k^2-p^2},
\eal
where $p=|\bp|$.
Following similar reasoning, we also obtain
\bal\label{transformationNonBravais}
&&\sum\limits_{\bR} e^{i\bk_B \cdot \bR}G_{\alpha\beta}(\bR\pm\mbf b) \nonumber\\
&&\qquad\qquad= \fr{1}{\mathcal A}\sum\limits_{\mbf G}g_{\alpha\beta}(\mbf G-\bk_B; 0)e^{\pm i\mbf b\cdot(\mbf G-\bk_B)}.
\eal

%%%%%%%%%%%%%%%%%%%%%%%%%%%%%%%%
\subsection{Green's function regularization and quantum fluctuations}\label{regularization}

In order to evaluate the right-hand side of \refeq{transformation}, special care has to be taken. While the left-hand side of \refeq{transformation} is finite and physically meaningful, the two terms on the right-hand side diverge individually --- only their difference is finite. %This divergence is unphysical and stems from the fact that we treat emitters as motionless, point-like particles.
An established technique to avoid such divergences is to take into account the quantum fluctuations of the particles \cite{Antezza2009,Antezza2009a}. %Here, we take into account the ground state fluctuations of the atoms in their harmonic trapping potential around their sites in the periodic lattice. 
These fluctuations `smear out' the divergent part of the Green's function at $\br=0$ over a finite volume, making it finite. In practice, this can be achieved by averaging the free-space Green's function with respect to the ground state wavefunction of a harmonically trapped atom \cite{Antezza2009,Antezza2009a}
\be\label{convolution}
G^*_{\alpha\beta}(\br) =\int d^3\bq \;G_{0,\alpha\beta}(\br-\bq) \mathcal |\psi_0(\bq)|^2,
\ee
where $\psi_0(\bq)$ is the ground state wavefunction of a quantum harmonic oscillator of frequency $\omega_\text{ho}=\hbar/(2ma_\text{ho}^2)$ given by
\bal\label{regularizer}
|\psi_0(\bq)|^2&=&\fr{1}{(\sqrt{2\pi}a_\text{ho})^3}e^{-q^2/2a_\text{ho}^2}.
\eal
Performing this integral with $\br=\mbf 0$ yields the following non-divergent, closed-form expression for the fluctuation-averaged Green's function at the source \cite{Antezza2009}
\bal\label{greensRegularizedAtSource}
G^*_{\alpha\beta}(\mbf 0)&=&\fr{k}{6\pi}\bigg[\bigg(\fr{\text{erfi}(ka_\text{ho}/\sqrt{2})-i}{e^{(ka_\text{ho})^2/2}}\bigg)\nonumber \\
&&\qquad\qquad\qquad- \fr{(-1/2)+(ka_\text{ho})^2}{(\pi/2)^{1/2} (ka_\text{ho})^3}\bigg]\delta_{\alpha\beta},\qquad
\eal
where $\text{erfi}(b)=2/\sqrt{\pi}\int_0^b dy \exp(y^2)$
is the imaginary error function.

The regularization techniques of Refs.~\cite{Antezza2009,Antezza2009a} can be specialized to the two-dimensional lattice to derive the fluctuation-averaged Weyl decomposition of the Green's function $g^*_{\alpha\beta}(\bp; 0)$ in the $z=0$ plane (see Appendix \ref{regularizationAppendix} for details of the derivation). The components of the resulting expression are given by
\bal\label{WeylGreensRegularized}
g^*_{xx}(\bp; 0) &=& (k^2-p_x^2)\mathcal{ I}_0,\nonumber\\
g^*_{yy} (\bp; 0)&=& (k^2-p_y^2)\mathcal{ I}_0,\nonumber\\
g^*_{zz} (\bp; 0)&=& (k^2\mathcal I_0-\mathcal I_2),\nonumber\\
g^*_{xy} (\bp; 0)&=& g^*_{yx}(\bp; 0) = -p_xp_y\mathcal{ I}_0\nonumber\\
g^*_{xz} (\bp; 0)&=& g^*_{zx}(\bp; 0)= 0,\nonumber\\
g^*_{yz} (\bp; 0)&=& g^*_{zy} (\bp; 0)= 0.
\eal \\
where
\bal\label{I0}
\mathcal I_0(p_x,p_y) =\mathcal C\; \frac{\pi e^{-a_\text{ho}^2\Lambda^2/2}}{\Lambda}\left[-i+\text{erfi}\left(\frac{a_\text{ho}\Lambda}{\sqrt{2}}\right)\right],\qquad\;
\eal
and
\bal\label{I2}
&&\mathcal{I}_2(p_x,p_y)=\mathcal C\;\bigg(-\fr{\sqrt{2\pi}}{a_\text{ho}}\nonumber\\
&&\qquad\qquad\quad+e^{-a_\text{ho}^2\Lambda^2/2}\pi \Lambda\left[-i+\text{erfi}\left(\frac{a_\text{ho}\Lambda}{\sqrt{2}}\right)\right]\bigg).\qquad\;
\eal
The functions $\mathcal{C}$ and $\Lambda$ depend on $p_x $ and $p_y$ and their explicit form is given by
\be
\mathcal C(p_x,p_y) = \fr{1}{2\pi k^2}e^{-a_\text{ho}^2(p_x^2+p_y^2)/2}.
\ee
and 
\bal
\Lambda(p_x,p_y) &=& (k^2-p_x^2-p_y^2)^{1/2}\geq 0.
\eal
%
%\be
%\mathcal I_0(p_x,p_y) = \fr{\pi e^{-a_\text{ho}^2(p_x^2+p_y^2+\Lambda^2)/2}}{2\pi k^2\Lambda}[-i+\text{erfi}(a_\text{ho}\Lambda/\sqrt{2})]
%\ee
%and
%\bal\label{I2}
%&&\mathcal{I}_2(p_x,p_y)=e^{-a_\text{ho}^2(p_x^2+p_y^2)/2}\bigg(-\fr{\sqrt{2\pi}}{a_\text{ho}}\nonumber\\
%&&\quad\qquad\quad+e^{-a_\text{ho}^2\Lambda^2/2}\pi \Lambda[-i+\text{erfi}(a_\text{ho}\Lambda/\sqrt{2}) ]\bigg),
%\eal
%with 
%\bal
%\Lambda(p_x,p_y) &=& (k^2-p_x^2-p_y^2)^{1/2},
%\eal
%where the square root with $\text{Im}(\Lambda)\geq 0$ and $\text{Re}(\Lambda)\geq 0$ is taken to preserve the causality of the Green's function \cite{Chew1995}. 
We note that the last two lines in \refeq{WeylGreensRegularized} are identically zero, since the polarization of the radiation emitted by a dipole is always parallel to the dipole orientation in the plane perpendicular to the dipole. 

%We these expressions we may write
%\bal
%&&\sum\limits_{\mathbf R\neq 0}e^{i\bk_B\cdot \mathbf R}G^*_{\alpha\beta}(\mathbf R)\nonumber\\
%&&\qquad\qquad=\left[\fr{1}{\mathcal{A}}\sum\limits_{\bG} g^*_{\alpha\beta}(\bG-\bk_B;0)-G^*_{\alpha\beta}(\mbf 0)\right].\qquad
%\eal

After substituting the regularized expressions from \refeq{greensRegularizedAtSource} and \refeq{WeylGreensRegularized} into \refeq{transformation} and \refeq{transformationNonBravais}, we may use \refeq{energiesBravais} and \refeq{chiBravais} (or \refeq{energiesNonBravais} and \refeq{chiNonBravais}) to obtain the eigenmodes of any Bravais (or non-Bravais) 2D lattice in the presence of atomic fluctuations.

Furthermore, the eigenmodes for point-like atoms (that are pinned to their lattice sites) can also be obtained through the simple modification of the fluctuation-averaged expressions for the Green's function.
In particular, it can be shown (see Ref.~\cite{Antezza2009} and Appendix \ref{cutoffIndependence}) that the expression
\be\label{cutoffIndependent}
e^{k^2a_\text{ho}^2/2}\fr{1}{\mathcal{A}}\sum_{\bG} g^*_{\alpha\beta}(\bG-\bk_B;0)-G^*_{\alpha\beta}(\mathbf 0)
\ee
approaches an $a_\text{ho}$-independent value as the limit ${a_\text{ho}\to 0}$ is taken. %(and the left-hand side of \refeq{transformation} is approached). 
Therefore, when $a_\text{ho}\ll 1/k$, the following approximation holds
\bal\label{TransformationRegularized}
&&\sum\limits_{\mathbf R\neq 0}e^{i\mathbf k_B\cdot \mathbf R}G_{\alpha\beta}(\mathbf R)\nonumber\\ 
&&\qquad\approx\fr{e^{k^2a_\text{ho}^2/2}}{\mathcal{A}}\sum\limits_{\bG} g^*_{\alpha\beta}(\bG-\bk_B;0)-G^*_{\alpha\beta}(\mbf 0).\quad
\eal
Using this approximation in \refeq{chiBravais} or \refeq{chiNonBravais}, we obtain the eigenmodes for arbitrary 2D lattice geometries made up of point-like atoms. 

\bigskip

The regularized expressions given above provide a straightforward, efficient and accurate way of calculating the eigenmodes of any 2D lattice composed of point-like or fluctuating atoms. Since we perform the summations in momentum space, only a few dozens of reciprocal lattice sites have to be included for full convergence. This contrasts sharply with performing summations in real space, where convergence often remains an issue even after tens of thousands of lattice sites have been included. To illustrate our formalism, we analyze below particular examples of 2D atomic lattices in free space and near planar metallic surfaces. 

%Using Eq.~\eqref{transformation}, \eqref{transformationNonBravais}, \eqref{greensRegularizedAtSource}, \eqref{WeylGreensRegularized} and \eqref{TransformationRegularized} we can now evaluta

% %%%%%%%
% \begin{figure}
%\centering
%\includegraphics[width=0.5 \textwidth]{./fig1_tri_free.png}
%\caption{
%\label{triangularFree}
%Triangular lattice in free space. (a) No B (b) B (c) Non-zero fluctuations $a_\text{ho}=0.4p$.} 
%\end{figure}
%%%%%%%%

%%%%%%%%%%%%%%%%%%%%%%%%%%%%%%%%
\section{Atomic lattices in free space}
%%%%%%%%%%%%%%%%%%%%%%%%%%%%%%%%

As an application of our formalism, we study 2D atomic lattices in free space. We focus on two examples -- a non-Bravais square lattice of three-level atoms and a triangular lattice of three-level atoms -- and discuss their topological properties. 

%%%%%%%
 \begin{figure}[h!]
\centering
\includegraphics[width=0.465 \textwidth]{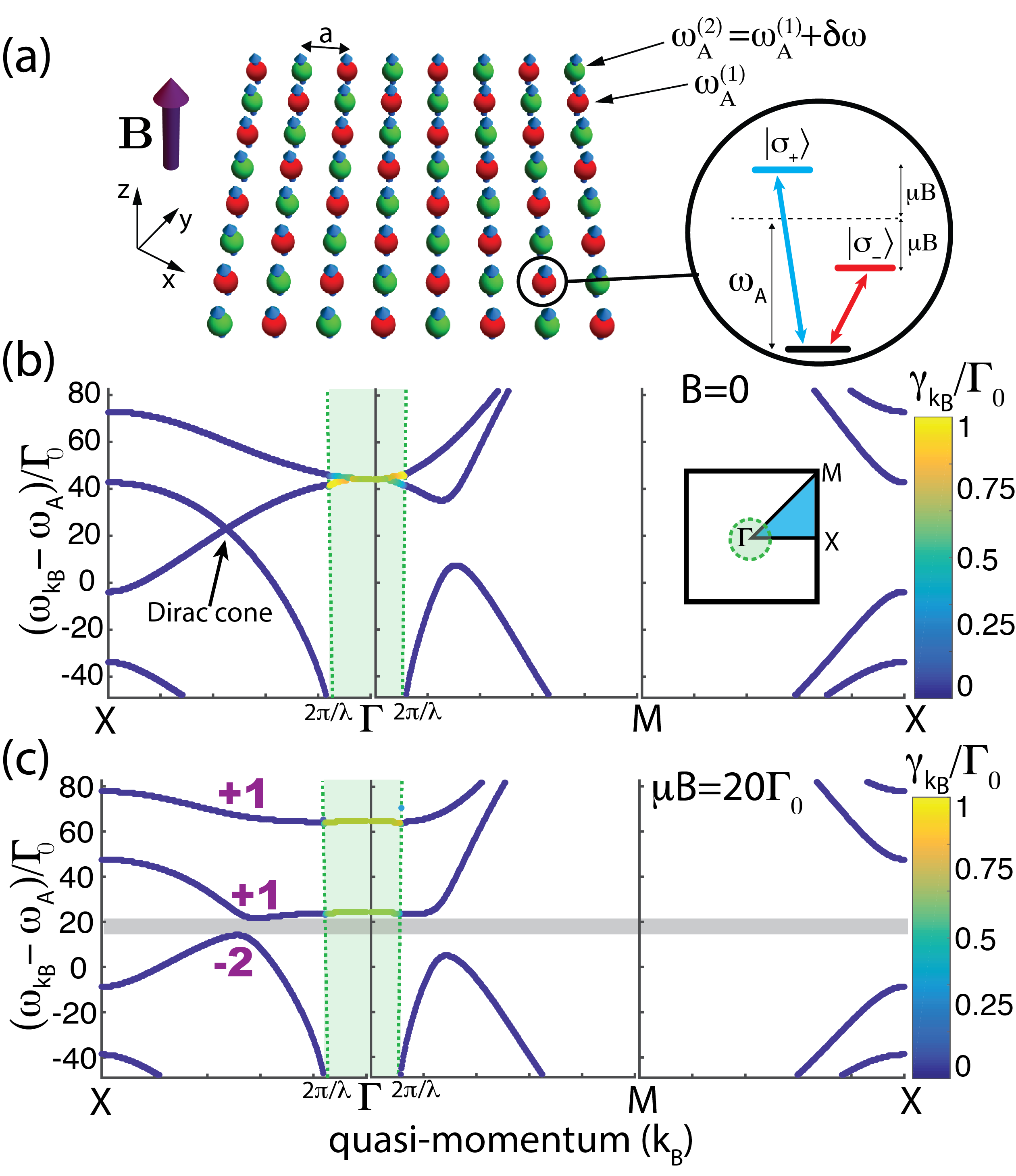}
\caption{
\label{NBSquareFree}
 (a) Non-Bravais square lattice of atoms in free space. Atoms marked in red (green) have resonant frequency $\omega_A^{(1)}$ ($\omega_A^{(2)} = \omega_A^{(1)}+\delta\omega$). Each atom has two excited states $\ket{\sigma_+}$ and $\ket{\sigma_-}$ that are Zeeman shifted by $\pm\mu B$ in the presence of an out-of-plane magnetic field $B$. (b) Part of the photonic band structure of the lattice for $B=0$. Green dashed lines mark the edges of the light cone. Modes that have quasi-momentum ${k_B<\omega_{\bk_B}}$ (green shaded region) can couple to free-space modes of the same momentum, making them short-lived. Modes that have quasi-momentum ${k_B>\omega_{\bk_B}}$ cannot couple to free-space modes and are long-lived. Decay rates of the modes are color coded. Bands are degenerate near the mid-point of the line joining the $\mbf X$ and $\mbf \Gamma$ points. (c) An out-of-plane magnetic field ($\mu B=20\Gamma_0$) opens a gap ($\Delta = 7\Gamma_0$) between the bands (grey shaded region). The bands acquire non-trivial Chern numbers (purple numbers).  Relevant parameters are $\lambda = 790\text{nm}$, $\delta\omega = 30\Gamma_0$, $\Gamma_0=2\pi\times 6\text{MHz}$ and $a=0.054\lambda$.}
\end{figure} 
%%%%%%%

%%%%%%%%%%%%%%%%%%%%%%%%%%%%%%%%
\subsection{Non-Bravais Square lattice of three-level atoms}\label{freeNBSquare}
%%%%%%%%%%%%%%%%%%%%%%%%%%%%%%%%

As our first example, we consider a non-Bravais square lattice of closely spaced ($a\approx \lambda/20$) atoms in free space as shown schematically in Fig.~\ref{NBSquareFree}(a). The properties of a non-Bravais honeycomb lattice in free space was studied previously in Ref.~\cite{Perczel2017}, where long-lived topological edge states were shown to exist on the system boundaries. Here, we show that a non-Bravais square lattice also supports long-lived topological edge excitations. These results demonstrate that these topological phenomena are not confined to any particular non-Bravais lattice geometry. Later, we will also see that these results stand in contrast with those obtained in Bravais lattices, where edge excitations are short-lived.  

The atomic non-Bravais square lattice lies in the $x$-$y$ plane with the quantization axis set along the $z$-axis and the interatomic spacing is $a$. Each atom is assumed to have two excited states ${\ket{\sigma_+}}$ and ${\ket{\sigma_-}}$, which can be excited by $\hat \sigma_+$ and $\hat \sigma_-$ polarized light, respectively. Note that the atoms could also have a transition to the $\ket{z}$ state, but from \refeq{freespaceGreensRealSpace} (see also \refeq{WeylGreensRegularized}) it follows that ${G_{xz}=G_{zx}=G_{yz}=G_{zy}=0}$ in the $x$-$y$ plane and, therefore, the $\sigma_\pm$ transitions are decoupled from the transition to the $\ket{z}$ state and we can ignore the latter. The lattice is assumed to consist of atoms of two different resonant transition frequencies $\omega_A^{(1)}$ (red atoms in Fig.~\ref{NBSquareFree}(a)) and ${\omega_A^{(2)} = \omega_A^{(1)} + \delta\omega}$ (green atoms), where $\delta\omega$ is a non-zero energy shift that may originate from having two different atomic species or a position-dependent Stark-shift. Here we assume that the atoms are point-like and their position is fixed.

First, we consider the atomic lattice in the absence of a magnetic field. After substituting \refeq{transformationNonBravais} and \refeq{TransformationRegularized} into \refeq{chiNonBravais}, we diagonalize the matrix in \refeq{energiesNonBravais} to obtain the complex eigenvalues $E_{\bk_B}=\omega_{\bk_B} - i\gamma_{\bk_B}$ for each Bloch vector $\bk_B$ inside the Brillouin zone. Fig.~\ref{NBSquareFree}(b) shows the resulting band structure along the lines joining the high symmetry points $\mbf \Gamma$, $\mbf M$ and $\mbf X$ inside the irreducible Brillouin zone (see inset of Fig.~\ref{NBSquareFree}(b)).

The decay rates of the modes ($\gamma_{\bk_B}$) are shown using a color code. The edges of the light cone are marked by green dashed lines at $k_B=2\pi/\lambda$. These lines correspond to free space modes propagating in the $x$-$y$ plane with maximal in-plane momentum $k_B=\omega_{\bk_B}/c$. All other free-space modes have an in-plane momentum component that satisfies $k_B<\omega_{\bk_B}/c$ (green shaded region in Fig.~\ref{NBSquareFree}(b)). The hybridized atom-photon modes of the atomic lattice with quasi-momentum ${k_B<\omega_{\bk_B}/c}$ can couple to free-space modes with matching momentum and energy, making these lattice modes short-lived. In contrast, lattice modes with quasi-momentum ${k_B>\omega_{\bk_B}/c}$ cannot couple to any of the free-space modes due to the momentum mismatch (since ${\left<k|k'\right>=\delta_{kk'}}$ in the momentum eigenbasis) \cite{Perczel2017,Asenjo-Garcia2017}. Therefore, lattice modes with $k_B>\omega_{\bk_B}/c$ are decoupled from free-space modes and do not decay when the lattice is infinite. 

This distinction between short-lived modes inside the light cone and long-lived modes outside the light cone is well-known in the literature of photonic crystal slabs, where periodic subwavelength dielectric structures are used to confine light in quasi-2D structures \cite{Joannopoulos2008}. We also note that even though the edges of the light cone indeed trace out a conical shape in $k$-space, the dashed green lines in Fig.~\ref{NBSquareFree}(b) appear vertical as we are only looking at a small energy range of a few linewidths around the atomic transition frequency and $\Gamma_0\ll \omega_A$ at optical frequencies.  

Due to the underlying symmetries of the lattice, the bands in Fig.~\ref{NBSquareFree}(b) form Dirac cones along the paths joining the $\mbf \Gamma$ point with the four $\mbf X$ points and there is a quadratic degeneracy at the $\mbf \Gamma$ point. These degeneracies arise due to the degeneracy of the $\ket{\sigma_+}$ and $\ket{\sigma_-}$ transitions in the absence of a magnetic field. When a magnetic field $\mbf B = B\hat z$ is switched on, the energy levels of the $\ket{\sigma_\pm}$ transitions are shifted by $\pm\mu B$ due to Zeeman splitting. %The effect of the magnetic field can be accounted for by introducing off-diagonal coupling terms of the form $V_\text{Zeeman}=i \mu B\ket{x}\bra{y} + \text{h.c}$ into the Hamiltonian. 
Breaking the degeneracy of the $\ket{\sigma_+}$ and $\ket{\sigma_-}$ transitions lifts the degeneracy of the bands and a complete band gap forms across the Brillouin zone (grey shaded band in Fig.~\ref{NBSquareFree}(c)). 

We investigate the topological character of the bands by calculating their Chern numbers \cite{Bernevig2013}, which are defined via the integral  
\be
C = \fr{1}{2\pi}\int \nabla_{\mbf k_B}\times \mathcal A (\mbf k_B)\cdot \text{d}\mbf k_B.
\ee 
The integral is performed over each band inside the irriducible Brillouin zone and the integrand is curl of the `Berry curvature', which is defined via
\be
\mathcal A(\mbf k_B) = \left<\psi_{\mbf k_B} \right| i \nabla_{\mbf k_B} \left| \psi_{\mbf k_B}\right>,
\ee 
where $\psi_{\mbf k_B}$ is the wavefunction of the Bloch mode with Bloch momentum $\mbf k_B$. We use the discretization method in Ref.~\cite{Fukui2005} to numerically obtain the Chern numbers for the bands. The resulting non-zero Chern numbers for the bands above and below the gap are shown in Fig.~\ref{NBSquareFree}(c). The four Dirac points between the two middle bands contribute +2 and -2 to the Chern numbers of the upper and lower bands respectively. At the same time, the quadratic degeneracy between the top two bands contributes +1 and -1 to the upper and lower bands respectively. Thus the three bands, from top to bottom, acquire Chern numbers +1, (+2-1)=+1 and -2. 

%%%%%%%
\begin{figure}
\centering
\includegraphics[width=0.45 \textwidth]{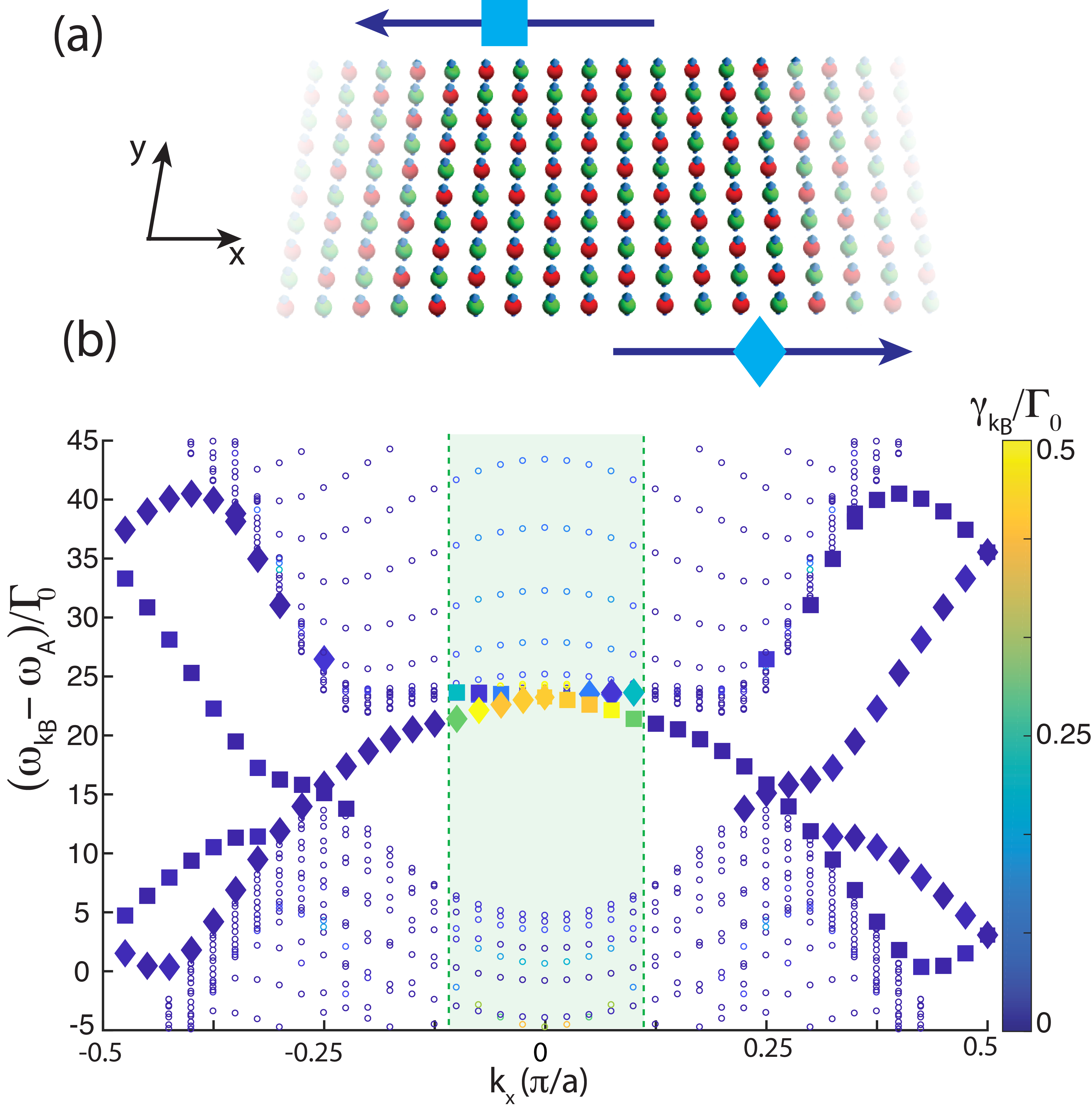}
\caption{
\label{NB_square_edge_states}
(a) Periodic strip of atoms in a non-Bravais square lattice. Each edge has edge modes propagating only in a single direction. (b) Modes that propagate on the upper (lower) edge are marked by squares (diamonds) in the band structure diagram. Extended bulk modes are marked with dots. There are two sets of edge modes on each boundary. Decay rates of modes are color coded. The edge modes cross the gap with quasi-momentum $k_B>\omega_{\bk_B}$ and thus are long-lived. Parameters are the same as in Fig.~\ref{NBSquareFree}(c). The spectrum was obtained for a 80x40 lattice of atoms with periodic boundary conditions along one direction.  A state is classified as an edge state on the (upper) lower boundary if the excitation probability on the top (bottom) four rows is at least 15 times the excitation probability on the bottom (top) four rows. }
\end{figure} 
%%%%%%%

\begin{figure}
\centering
\includegraphics[width=0.45 \textwidth]{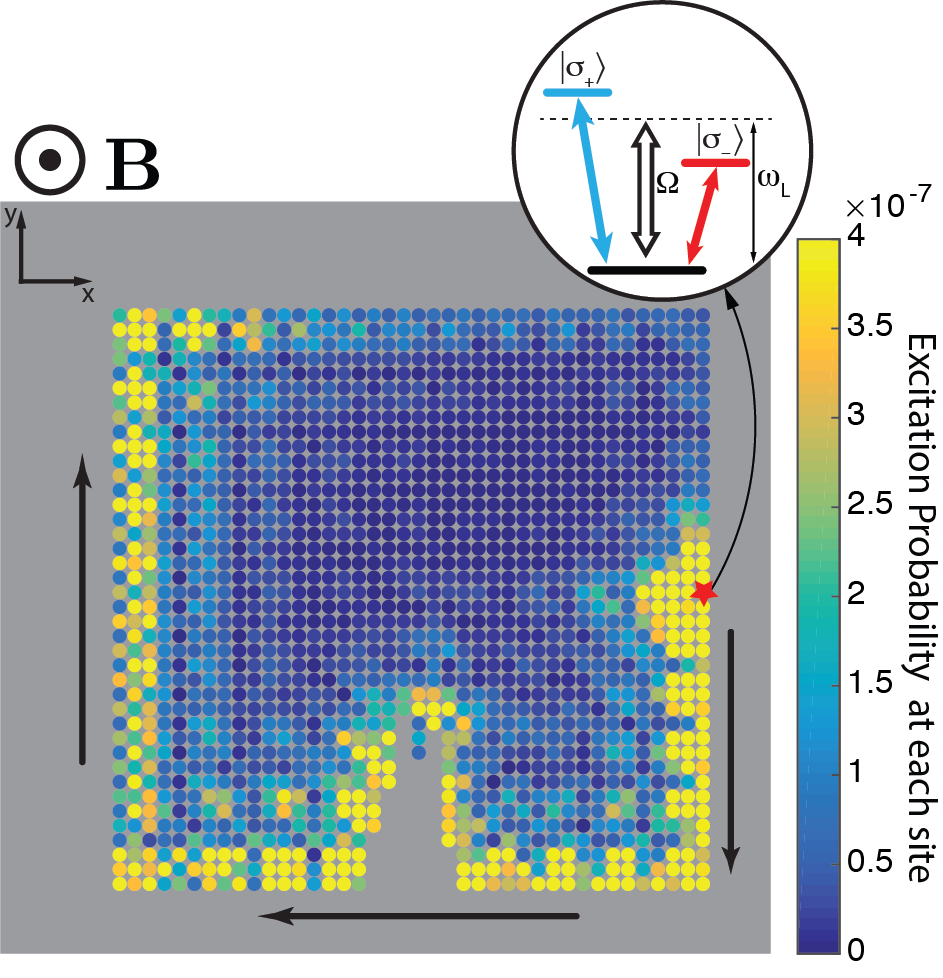}
\caption{
\label{NB_square_time}
Snapshot of the time evolution (at $t=11.1\Gamma_0^{-1}$) of a finite non-Bravais square lattice in free space with an irregular defect. An atom on the edge (marked by the red star) is driven by a laser (see inset). The color code depicts the excitation probability ${|\bra{\psi(t)}\sigma^i_{+}\big>|^2+|\bra{\psi(t)}\sigma^i_-\big>|^2}$ at each site. Approximately 96\% of the emitted excitation is coupled into the forward direction. Emission into bulk modes and the backward edge modes is suppressed. The excitation goes around the corners and routes around the large, irregular defect, which demonstrates the robustness of the topological excitation to disorder. Relevant parameters are $N=1576$, $\lambda=790\text{nm}$, $\delta\omega = 30\Gamma_0$, ${\Gamma_0= 2\pi\times6\text{MHz}}$, $a=0.054\lambda$ and $\mu B=20\Gamma_0$. The strength of the drive is $\Omega=0.1\Gamma_0$ and the driving frequency is ${\omega_\text{L}=\omega_A+18\Gamma_0}$. The driving laser is adiabatically switched on with a Gaussian profile $\Omega(t)=\Omega\exp(-[t-4.5\Gamma_0^{-1}]^2/[1.35\Gamma_0^{-2}])$ for $t<4.5\Gamma_0^{-1}$.}
\end{figure}

Band gaps between topological bands are associated with edge states \cite{Bernevig2013}. We investigate the spectrum of edge states by finding the eigenmodes of a periodic strip of the atomic lattice (see \reffig{NB_square_edge_states}(a)). The calculation proceeds by defining an MxN lattice of atoms in real space, where the lattice has periodic boundary conditions along the first direction (\reffig{NB_square_edge_states}(a)). The interactions between atoms is calculated using \refeq{Hamiltonian} and the range of the interactions is truncated after M/2 sites \footnote{Since the interactions are long range $(\sim 1/r)$, assuming periodic boundary conditions along one direction requires the truncation of the range of interaction. This leads to small deviations in the decay rates, potentially making them negative. This is an artifact of our numerical method with no physical significance. When the edge modes are obtained for a lattice of atoms that is finite in all directions (see Fig.~\ref{NB_square_time}), no such truncation is necessary and all decay rates are non-negative.}. The wavefunctions of the resulting eigenmodes are then Fourier-analyzed to find $k_x$, the component of the Bloch quasi-momentum along the x-axis, associated with each mode.   

\reffig{NB_square_edge_states}(b) shows the spectrum of  $H_{\rm eff}$ for such a configuration, where the eigenergies are plotted versus $k_x$. Edge modes on the lower and upper boundaries of the strip are marked by diamonds and squares respectively. The edge states cross the gap connecting the two bulk bands. Since the sum of the Chern numbers above and below the band gap is +2 and -2 respectively, there are two sets of edge modes on each boundary. Note that both sets of edge modes on the lower boundary carry energy to the right, whereas both sets of modes on the top boundary carry energy to the left. Since energy flow is unidirectional on each boundary, these edge states carry energy forward without backscattering. Crucially, we find that the edge modes cross the gap outside the light cone with quasi-momenta $k_B>\omega_{\bk_B}/c$. Therefore, these edge states do not couple to free-space modes due to the momentum mismatch, making them long-lived with decay rates much smaller than $\Gamma_0/2$. This suppression of losses is the key result, which leads to long-lived edge excitations. Such edge excitations can carry energy around the boundaries of the lattice with minimal losses.

%%%%%%%
%%%%%%%

\reffig{NB_square_time} shows a snapshot of the excitation probabilities during the no-jump time evolution of the system when a single atom is continuously excited by a laser on the boundaries of the lattice. If the driving field is weak, the dynamics are essentially captured by the single excitation description introduced above. We add to \refeq{NBwavefunction} the ground state component $+c_g\ket{g_*}$ of the driven atom and $\ket{g_*}$ is coupled to the excited states $\ket{\sigma_{+,*}}$ and $\ket{\sigma_{-,*}}$ of the atom by adding the driving terms ${\Omega(t)\left(\ket{\sigma_{+,*}}\bra{g_*} + \ket{\sigma_{-,*}}\bra{g_*} + \text{h.c.}\right)}$ to the effective Hamiltonian $H_\text{eff}$. We obtain the time-evolved wavefunction at time $t$ by numerically finding ${\ket{\psi(t)}=\exp \left(-iH_\text{eff}/\hbar t\right)}\ket{\psi(0)}$, where the excitation is initially in the ground state of the driven atom ($\left| \left<g_*|\psi(0)\right>\right|^2=1$). The laser is resonant with the edge states inside the band gap and is switched on adiabatically to avoid exciting non-resonant modes. We find that approximately 96\% of the excitation emitted by the atom is coupled into the edge modes carrying energy in the clockwise direction, while coupling into modes circulating anti-clockwise and into the bulk modes is strongly suppressed due to the gap and topology. Given the absence of channels for backscattering and highly suppressed coupling to free space modes, the excitation routes around corners with approximately 97\% efficiency as well as around the irregular defect with approximately 86\% efficiency. We emphasize that these results illustrate the key point that topological quantum optical systems are protected against both losses into free space and large defects in the lattice. We also note that qualitatively similar results were obtained in the non-Bravais honeycomb lattice in Ref.~\cite{Perczel2017}.

%%%%%%%%%%%%%%%%%%%%%%%%%%%%%%%%
\subsection{Triangular lattice of three-level atoms}\label{freeTriangular}
%%%%%%%%%%%%%%%%%%%%%%%%%%%%%%%%

\begin{figure}[h!]
\centering
\includegraphics[width=0.5 \textwidth]{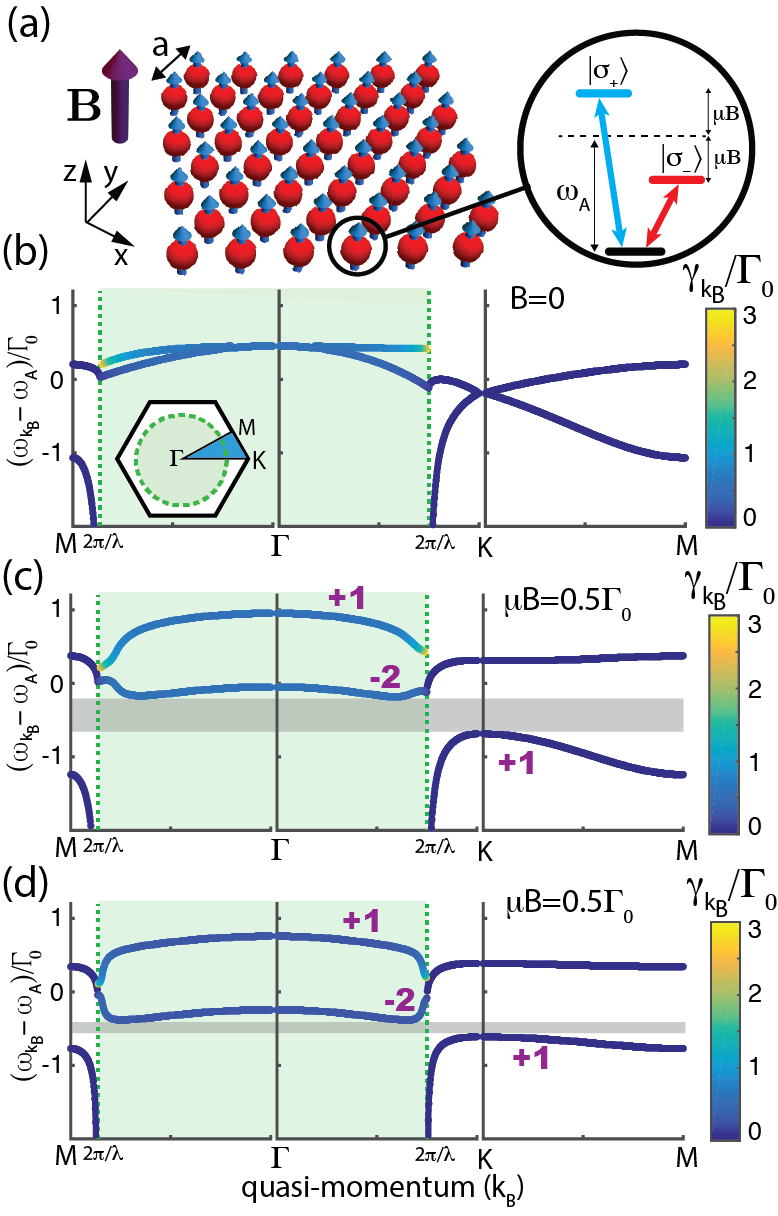}
\caption{
\label{triangularFree}
(a) Triangular lattice of atoms with interatomic spacing $a$ in free space. Each atom has two excited states $\ket{\sigma_+}$ and $\ket{\sigma_-}$ with Zeeman-splitting $\pm\mu B$ due to an out-of-plane magnetic field $B$. (b) Photonic band structure for $B=0$ and point-like atoms pinned to their lattice sites. Green dashed lines mark the light cone edges. Modes with quasi-momentum ${k_B<\omega_{\bk_B}}$ (green shaded region) couple to free-space, making them short-lived. Modes with quasi-momentum ${k_B>\omega_{\bk_B}}$ are long-lived. Decay rates of the modes are color coded. Bands are degenerate at the $\mbf \Gamma$ and $\mbf K$ points. (c) A transverse magnetic field ($\mu B=0.5\Gamma_0$) opens a small gap ($\Delta = 0.6\Gamma_0$) between the bands (grey shaded region). The bands acquire non-trivial Chern numbers (purple numbers). (d) When the atoms fluctuate around the lattice sites, the size of the gap is reduced. Here the amplitude of fluctuations is $a_\text{ho}=0.4a$. Relevant parameters for all three plots are $\lambda = 790\text{nm}$, $\Gamma_0=2\pi\times 6\text{MHz}$ and $a=\lambda/2$. } 
\end{figure}

Topological photonic bands in atomic lattices can also be obtained when $a\approx \lambda/2$ --- a trapping regime that is routinely explored in cold atom laboratories worldwide \cite{Bloch2005}. Fig.~\ref{triangularFree}(a) shows a triangular Bravais lattice of atoms in the $x$-$y$ plane in free space with $a=\lambda/2$ and quantization axis along the $z$-axis. As in the case of the non-Bravais square lattice, each atom is assumed to have two excited states $\ket{\sigma_+}$ and $\ket{\sigma_-}$.
 %%%%%%%
%%%%%%%

Initially, we assume point-like atoms in zero magnetic field. After substituting \refeq{TransformationRegularized} into \refeq{chiBravais}, we diagonalize the matrix in \refeq{energiesBravais} to obtain the complex eigenvalues. Fig.~\ref{triangularFree}(b) shows the band structure along lines joining the high symmetry points $\mbf \Gamma$, $\mbf K$ and $\mbf M$ of the Brillouin zone (see inset of Fig.~\ref{triangularFree}(b)). 

The decay rates of the modes $\gamma_{\bk_B}$ are encoded using a color code and the light cone edges are marked by green dashed lines. Given the $a=\lambda/2$ spacing of the atoms, the boundaries of the irreducible Brillouin zone are close to the light cone edges. Therefore, the modes are decoupled from free-space modes only in a small region near the perimeter of the irreducible Brillouin zone. 

The bands in Fig.~\ref{triangularFree}(b) form a Dirac cone at the $\bK$ point and a quadratic degeneracy at the $\mbf \Gamma$ point. When a magnetic field $\mbf B = B\hat z$ is switched on, the degeneracies are lifted and a small gap forms across the Brillouin zone (shaded band in Fig.~\ref{triangularFree}(c)). The bands acquire non-zero Chern numbers (purple numbers) indicating the topological nature of the bands.

Fig.~\ref{triangularFree}(d) shows the band structure in the presence of a magnetic field when the fluctuations of the atoms are taken into account. The plot is obtained by substituting the regularized expressions from \refeq{greensRegularizedAtSource} and \refeq{WeylGreensRegularized} into \refeq{transformation} and diagonalizing \refeq{energiesBravais} %and \refeq{chiBravais}
with $a=\lambda/2$ and $a_\text{ho}=0.4a$. Since fluctuations smear out the relative phases between sites, the band gap becomes smaller relative to the case when atoms were assumed to be point-like. For even larger fluctuations the gap closes. We note that the band structure is not significantly affected by fluctuations as long as $a_\text{ho}\lesssim 0.15 a$.

The presence of the gap between the topologically non-trivial bands implies that for a finite lattice, edge states appear on the system boundaries. However, since the boundaries of the light cone are very close to the boundaries of the irreducible Brillouin zone, all edge states will fall within the light cone. Therefore, all edge states couple to the free-space modes, making them short-lived. This makes energy transfer with these edge states impractical. The spacing of the atoms can be reduced slightly to increase the size of the Brillouin zone and establish larger areas that are outside the light cone. However, as the spacing between atoms is reduced, the gap size gradually decreases and closes before the edge states cross the edges of the light cone. This stands in contrast with the case of a honeycomb lattice of atoms, where the gap size increases as $\sim \! 1/a^3$ with decreasing interatomic distance \cite{Perczel2017}.   

%Closing of the gap with decreasing interatomic spacing is due the fact that in this regime the coupling between atoms is mediated by long-range photons with $k=2\pi/\lambda$, which only couple to Bloch modes of the lattice with $k_B=\pi/a\approx \pi/(\lambda/2)$.

Qualitatively similar results can be obtained in a square Bravais lattice of atoms with $a\approx \lambda/2$ and identical atomic level structure. In this case, quadratic degeneracies form  both at the $\mbf \Gamma$ and $\mbf M$ points. A magnetic field lifts the degeneracies, opening a gap. The size of the gap varies with interatomic spacing and is non-negligible ($\Delta \sim 0.5\Gamma_0$) only when $a\approx \lambda/2$. The maximum gap size is generally smaller in the square lattice than for a triangular lattice with similar parameters, as the Brillouin zone of the square lattice is less circular than that of the triangular lattice \cite{Karzig2015}.

%%%%%%%%
%\begin{figure}
%\centering
%\includegraphics[width=.45 \textwidth]{./fig6_Greens_channels.png}
%\caption{
%\label{ScatteredGreens}
%Green's function components in the vicinity of a metal surface.} 
%\end{figure} 
%%%%%%%%

%%%%%%%%%%%%%%%%%%%%%%%%%%%%%%%%
\section{Atomic lattices near a metalic planar surface}\label{planarGreens}
%%%%%%%%%%%%%%%%%%%%%%%%%%%%%%%%

%NEED TO REPLACE $\Gamma_0\to\Gamma_\text{sc}$ in Hamiltonian!

%The techniques we have developed so far can also be used to analyze emitter lattices near metallic surfaces. 
The study of dipole-like atomic emitters near metallic surfaces has received significant attention lately as part of recent research efforts into plasmonics \cite{Ebbesen1998,Chang2006,GarciadeAbajo2007,Nikitin2010,Kildishev2013,Tame2013,Delga2014,High2015}. The technique developed so far can be extended in a straightforward way to describe atomic lattices in the vicinity of planar metal surfaces as long as the plane of the atomic lattice is parallel to all surfaces, thus ensuring that translational invariance is preserved in the $x$-$y$ plane. Here, we analyze atoms placed near a single metal-dielectric interface. In Appendix \ref{greensNearFlatSurface}, we also discuss how our method generalizes to describe atomic lattices in the presence of arbitrarily layered planar media. 

We assume that the plane of the atomic lattice lies at $z=0$ and the metal-dielectric interface is located at a distance $h$  below the dipoles at $z=-h$. The eigenmodes of the atomic lattice are then obtained form \refeq{energiesBravais} and \refeq{energiesNonBravais} by substituting $G_{\alpha\beta}$ into \refeq{chiBravais} and \refeq{chiNonBravais}, where $G_{\alpha\beta}$ is the solution of \refeq{greensEquation} with the following spatially inhomogeneous and frequency-dependent permittivity
\be
\vare(\omega,\br)=
\begin{cases}
\vare_d, & \text{if}\ z>-h\\
\vare_m(\omega)<0, & \text{if}\ z<-h
\end{cases},
\ee
where $\vare_d$ is the permittivity of the dielectric and $\vare_m(\omega)$ is the frequency-dependent permittivity of the metal. In the presence of a planar interface, there is no known closed-form solution for $G_{\alpha\beta}$ in position space. In contrast, it is possible to exactly solve for the Weyl decompositon $g_{\alpha\beta}(\bp;z)$ in momentum space in the presence of arbitrarily layered planar media (see Appendix \ref{greensNearFlatSurface} for more details), which allows us to efficiently perform the band structure calculations using the expressions \refeq{transformation} and \refeq{transformationNonBravais} \footnote{We note that using the exact Green's function in momentum space has the other advantage of allowing us to account for the interaction of the atoms not only with surface plasmons, but also with other types of surface waves, which are collectively known as ``creeping waves'' or ``quasi-cylindrical waves'' \cite{Lalanne2006,Liu2008,Lalanne2009,Nikitin2010}}. %While at optical frequencies creeping waves decay fast far away from the source and are often neglected, their contribution to the radiation field becomes significant when the distance to the source is comparable to the wavelength \cite{Nikitin2010}. Indeed, separating out and considering only the plasmonic component of the surface excitations can lead to unphysical predictions in the near field. For example, the purely plasmonic response of a metal is predicted to diverge at the location of the dipole {\it regardless} of the distance of the emitter to the surface \cite{Archambault2009}. When both the creeping wave and plasmonic components are included, no such divergences arise.  

For illustration, below we analyze two examples of atomic lattices near plasmonic surfaces -- a square lattice of four-level atoms featuring a non-topological gap and a triangular lattice of three-level atoms giving rise to topological gaps.

%%%%%%%
\begin{figure}
\centering
\includegraphics[width=0.5 \textwidth]{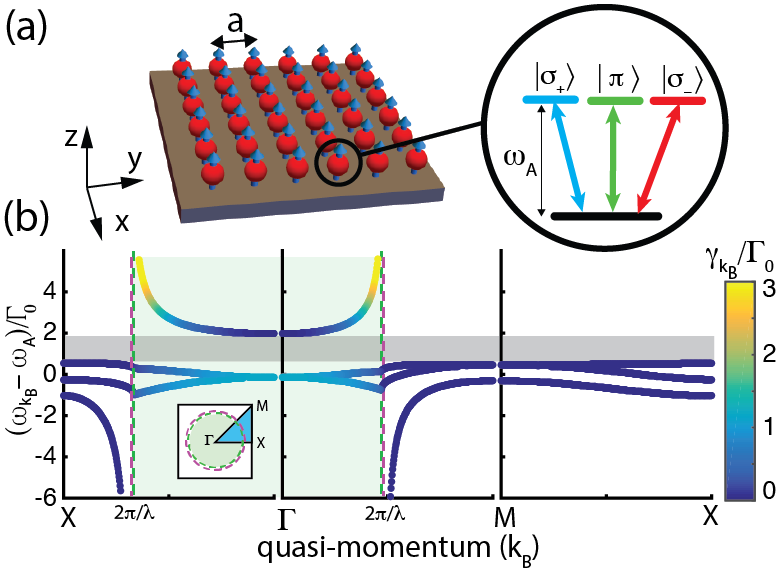}
\caption{
\label{square_plasmon}
(a) Square lattice of atoms with interatomic spacing $a$ and distance $h$ from a silver surface. Each atom has three transitions to the $\ket{\sigma_+}$, $\ket{\pi}$ and $\ket{\sigma_-}$ states. (b) Band structure of the lattice. Green dashed lines mark the edges of the free-space light cone (green shaded region), purple dashed lines show the plasmonic dispersion of the metal surface. Decay rate of the modes is color coded. Due to the strong coupling of the $\ket{\pi}$ state to the plasmons, a non-topological gap opens in the spectrum (grey shaded region). Relevant parameters are $\lambda = 737\text{nm}$, $\Gamma_0=2\pi\times 0.95\text{MHz}$, $a=\lambda/3$ and $h=\lambda/10$.}
\end{figure} 
%%%%%%%

%%%%%%%%%%%%%%%%%%%%%%%%%%%%%%%%
\subsection{Square lattice of four-level atoms}
%%%%%%%%%%%%%%%%%%%%%%%%%%%%%%%%

As our first example, we consider a square lattice of atoms in free space ($\vare_d =1$) near a silver surface as shown in \reffig{square_plasmon}(a). Each atom is assumed to have three excited states $\ket{\sigma_+}$, $\ket{\sigma_-}$ and $\ket{\pi}$, which are excited by $\hat \sigma_+$, $\hat \sigma_-$ and $\hat z$ polarized light. The atoms are assumed to be very close to the surface $h\ll \lambda$, such that coupling to plasmons is strong at optical frequencies. For the frequency-dependent dielectric permittivity of metal we use the material properties reported for single-crystal silver in Ref.~\cite{High2015}, which were experimentally shown to give rise to propagation distances on the order of 150-200$\lambda$ for plasmons at optical frequencies. Therefore, we disregard the imaginary part of the permittivity with the understanding that the propagation distance of the hybrid atom-plasmon modes of the lattice is eventually limited by ohmic losses. For a discussion of the notion of band gaps in the presence of losses see Ref.~\cite{Deutsch1995}.

 \reffig{square_plasmon}(b) shows the band structure of the atomic lattice near the metal surface with $a=\lambda/3$ and $h=\lambda/10$. Due to the proximity of the metal surface, the in-plane components of the electromagnetic field are small. Therefore, the bare resonant energy $\omega_A$ of the $\ket{\sigma_+}$ and $\ket{\sigma_-}$ transitions of the individual atoms is not strongly affected by the presence of the other atoms, resulting in the two nearly flat bands near $\omega_A$. In contrast, the $\ket{\pi}$ transition couples strongly to plasmons, which are predominantly polarized along $\hat z$. The strong interaction splits the $\hat z$ polarized band into two bands (top and bottom band in  \reffig{square_plasmon}(b)), resulting in a (non-topological) band gap. At energies inside the band gap no plasmonic modes can propagate.  

The edges of the light cone are marked by green dashed lines in \reffig{square_plasmon}(b). For frequencies far away from the plasma frequency, the plasmonic modes fall just outside the edges of the light cone. These plasmonic modes are marked by magenta dashed lines in  \reffig{square_plasmon}(b). Modes within the light cone couple strongly to free space modes, whereas modes outside the light cone do not decay.

 %%%%%%%
\begin{figure}
\centering
\includegraphics[width=0.5 \textwidth]{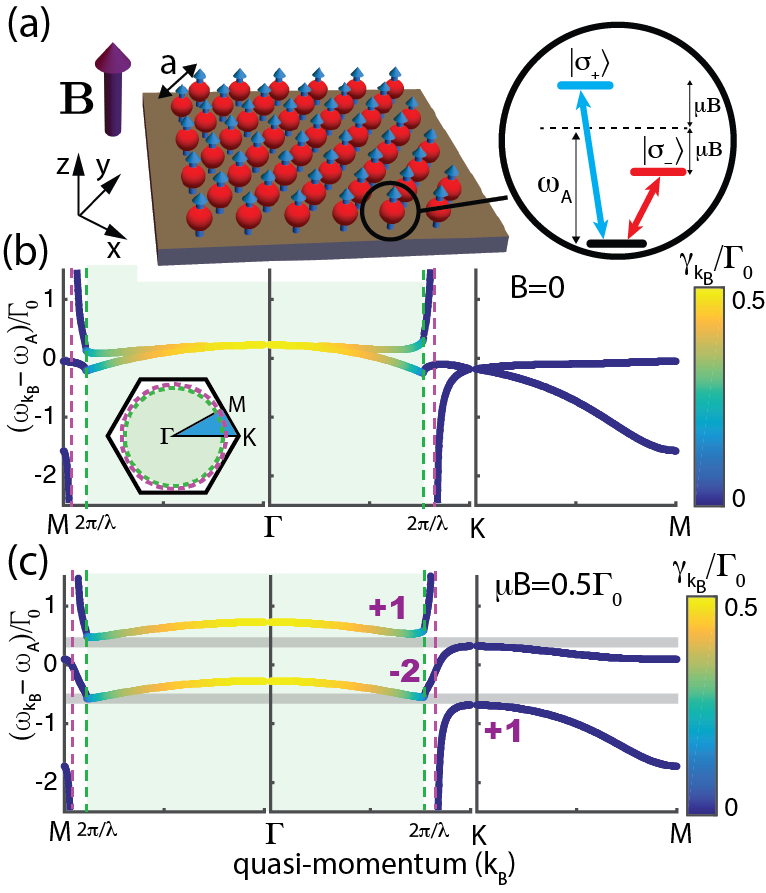}
\caption{
\label{triangular_plasmon}
(a) Triangular lattice of atoms with transitions to the $\ket{\sigma_+}$ and $\ket{\sigma_-}$ states near a metal surface. (b) Band structure of the lattice for $B=0$. Green dashed lines mark the edges of the free-space light cone (green shaded region), purple dashed lines show the plasmonic dispersion of the metal surface. Decay rate of the modes is color coded. Bands are degenerate at the $\mbf \Gamma$ and $\mbf K$ points. (c) A transverse magnetic field ($\mu B = 0.5\Gamma_0$) opens two small gaps between bands that have non-trivial Chern numbers. The relevant parameters are $\lambda = 437\text{nm}$, $\Gamma_0=2\pi\times 0.95\text{MHz}$, $a=\lambda/1.95$ and $h=\lambda/15$.}
\end{figure} 
%%%%%%%

%%%%%%%%%%%%%%%%%%%%%%%%%%%%%%%%
\subsection{Triangular lattice of three-level atoms}
%%%%%%%%%%%%%%%%%%%%%%%%%%%%%%%%

Hybridized atom-plasmon-photon bands may also have topological character in the proximity of the metal surface. \reffig{triangular_plasmon} shows a triangular lattice of three-level atoms with V-level structure near a metal surface, where the transition to the $\ket{\pi}$ state is not included since it decouples completely from the transition to the $\ket{\sigma_+}$ and $\ket{\sigma_{-}}$ states (see discussion in Section \ref{freeNBSquare}). The $\ket{\sigma_+}$ and $\ket{\sigma_-}$ transitions are assumed to be near-UV ($\lambda = 437$nm). Such transitions are favorable, since at higher frequencies the plasmons get more tightly confined to the metal surface and the in-plane components of the plasmonic fields increase, making interactions with the $\ket{\sigma_\pm}$ transitions stronger. \reffig{triangular_plasmon}(b) shows the band structure in the absence of a magnetic field. A quadratic degeneracy forms at the $\mbf \Gamma$ point and a Dirac point is found at the $\mbf K$ point, just as in free space. Applying a magnetic field lifts the degeneracies and opens up two small gaps in the spectrum (grey shaded bands in \reffig{triangular_plasmon}(c)) and the bands acquire non-trivial Chern numbers (purple numbers). The bands remain topological even if atoms with longer transition wavelength are used. However, in this case the interactions will be weaker, making the `avoided crossings' of bands smaller and the gap disappears. Note that the band structures in \reffig{triangular_plasmon}(b) and (c) are qualitatively similar to the band structures obtained for a triangular lattice of atoms in free space (\reffig{triangularFree}(b) and (c)). The key difference is that in \reffig{triangular_plasmon} the uppermost band asymptotically approaches the unperturbed plasmonic modes of the metal surface (purple dashed lines) for ${\omega_{\bk_B}}\gg \omega_A$, whereas in \reffig{triangularFree} the decay rate of the uppermost band diverges as the edges of the light cone are approached, effectively dissolving the band \cite{Perczel2017}.

%%%%%%%%%%%%%%%%%%%%%%%%%%%%%%%%
\section{Conclusions}\label{conclusion}
%%%%%%%%%%%%%%%%%%%%%%%%%%%%%%%%

We have described a general method for calculating the photonic band structure for infinite two-dimensional atomic lattices in free space and near planar surfaces with arbitrary Bravais and non-Bravais geometries. This method takes into account the full radiation pattern emitted by individual atoms, which gives rise to long-range interactions that scale as $\sim 1/r$. By performing the required summations in momentum space rather than in real space, calculation of collective energy shifts and decay rates can be performed efficiently. This method makes it possible to account for atomic position fluctuations and can be extended to describe atomic lattices near metallic surfaces.

As applications of our method, we studied non-Bravais square lattices and triangular Bravais lattices in free space and investigated their topological properties, including topological edge states in the band gap. We also obtained the band structure of an atomic square lattice near metallic surfaces and studied the topological bands that arise when a triangular lattice of atoms is placed near the metal surface. Given the generality of our method, we expect that it will pave the way for further studies of two-dimensional atomic lattices both in free space and near planar surfaces and will facilitate finding experimentally accessible parameter regimes for the realization of topological quantum optics and other intriguing phenomena.  

%%%%%%%%%%%%%%%%%%%%%%%%%%%%%%%%
\section{Acknowledgements}\label{acknowledgements}
%%%%%%%%%%%%%%%%%%%%%%%%%%%%%%%% 

We thank Dominik Wild, Ephraim Shahmoon, Alex High and J. Taylor for illuminating discussions. We acknowledge funding from the MIT-Harvard CUA, NSF, AFOSR and MURI. JP acknowledges support from the Hungary Initiative Foundation. JB acknowledges funding from the Carlsberg Foundation, the Qubiz - Quantum Innovation Center, and VILLUM FONDEN via the QMATH Centre of Excellence (Grant No. 10059). DEC acknowledges support from the MINECO ``Severo Ochoa'' Program (SEV-2015-0522), MINECO Grant CANS, Fundacio Privada Cellex, CERCA Programme / Generalitat de Catalunya, and ERC Starting Grant FOQAL. HP acknowledges funding from NSF though a grant for ITAMP. Work at Innsbruck is supported by SFB FOQUS of the Austrian Science Fund, and ERC Synergy Grant UQUAM.

\section*{Appendix}

%%%%%%%%%%%%%%%%%%%%%%%%%%%%%%%%
%%%%%%%%%%%%%%%%%%%%%%%%%%%%%%%%
\appendix
%%%%%%%%%%%%%%%%%%%%%%%%%%%%%%%%
%%%%%%%%%%%%%%%%%%%%%%%%%%%%%%%%

%%%%%%%%%%%%%%%%%%%%%%%%%%%%%%%%
\section{Green's function expressions}\label{GreensAppendix}
%%%%%%%%%%%%%%%%%%%%%%%%%%%%%%%%

In free space the permittivity is $\vare = 1$ and \refeq{greensEquation} can be solved using Fourier techniques to obtain the following momentum-space representation of the dyadic Green's function \cite{Chew1995}
\be\label{freespaceGreens}
G_{0,\alpha\beta}(\br) =\int \fr{d^3\bp}{(2\pi)^3}e^{i\bp\cdot \br}\fr{1}{k^2}\fr{k^2\delta_{\alpha\beta}-p_\alpha p_\beta}{k^2-p^2},
\ee
where $\bp=p_x\hat x+p_y\hat y+p_z\hat z$, $p=|\bp|$ and $k=\omega/c$. \refeq{greensEquation} also has a closed-form solution in position space \cite{Chew1995} given by
\be\label{ScalarGreensDerivative}
 G_{0,\alpha\beta}(\br)=-\left(\fr{1}{k^2}\pa_\alpha\pa_\beta+\delta_{\alpha\beta}\right)\fr{e^{ikr}}{4\pi r}.
\ee
After evaluating the derivatives \cite{Frahm1983,Franklin2010}, we obtain the following Green's function components \cite{Dung1998,Morice1995}
\bal\label{freespaceGreensRealSpaceAppendix}
&&G_{0,\alpha\beta}(\br)=-\fr{e^{ikr}}{4\pi r}\bigg[\bigg( 1+\fr{i}{kr}-\fr{1}{(kr)^2} \bigg)\delta_{\alpha\beta}\qquad\nonumber\\
&&\qquad+\bigg(-1-\fr{3i}{kr}+\fr{3}{(kr)^2}\bigg) \fr{x_\alpha x_\beta}{r^2}  \bigg]+\fr{\delta_{\alpha\beta}\delta^{(3)}(\br)}{3k^2}.\qquad
\eal

%%%%%%%%%%%%%%%%%%%%%%%%%%%%%%%%
\section{Green's function regularization}\label{regularizationAppendix}
%%%%%%%%%%%%%%%%%%%%%%%%%%%%%%%%

We regularize the Green's function by inserting a Gaussian momentum cut-off of the form \cite{Antezza2009}
\be
\tilde {\mathcal R}(a_\text{ho},\bp) = e^{-a_\text{ho}^2p^2/2}
\ee
into the Fourier integral of the Green's function (Eq.~\ref{freespaceGreens}), which yields
\be\label{freespaceGreensRegularizedApp}
G^*_{0,\alpha\beta}(\br) =\int \fr{d^3\bp}{(2\pi)^3}e^{i\bp\cdot \br}\fr{k^2\delta_{\alpha\beta}-p_\alpha p_\beta}{k^2(k^2-p^2)}e^{-a_\text{ho}^2p^2/2}.
\ee 
The momentum cut-off removes the high-frequency contributions that are associated with momenta $p\gg1/a_\text{ho}$. \refeq{freespaceGreensRegularizedApp} is a convolution between the Green's function and the momentum cut-off and, therefore, by the convolution theorem we obtain
\be\label{convolution}
G^*_{0,\alpha\beta}(\br) =\int d^3\bq \;G_{0,\alpha\beta}(\br-\bq) \mathcal R(a_\text{ho},\bq),
\ee
where $\mathcal R(a_\text{ho},\bq)$ is the Fourier transform of the momentum cut-off function $\tilde {\mathcal R}(a_\text{ho},\bp)$ given by
\bal\label{regularizerApp}
\mathcal R (a_\text{ho},\mbf{q})&=&\int \fr{d^3\bp}{(2\pi)^3} e^{i\bp\cdot \mbf{q}}e^{-a^2_\text{ho}p^2/2}\nonumber\\
&=&\fr{1}{(\sqrt{2\pi}a_\text{ho})^3}e^{-q^2/2a_\text{ho}^2}\nonumber \\
&=&|\psi_0(\bq)|^2,
\eal
where $\psi_0(\bq)$ is the ground state wavefunction of a quantum harmonic oscillator of frequency $\omega_\text{ho}=\hbar/(2ma_\text{ho}^2)$. \refeq{convolution} thus represents the averaging over the ground state fluctuations of a harmonically trapped atom, where $a_\text{ho}$ is the amplitude of the ground state fluctuations \cite{Antezza2009}. Substituting \refeq{freespaceGreensRealSpace} into  \refeq{convolution}, setting $\br=\mbf 0$ and evaluating the integral gives Eq.~\eqref{greensRegularizedAtSource}. 

In order to find the fluctuation-averaged Weyl decomposition of the free-space Green's function in the plane of the atoms ($z=0$), we need to evaluate the following expression 
\bal\label{pzRegularizedApp}
g^*_{\alpha\beta}(\bp; 0)=\int \fr{dp_z}{2\pi}\fr{k^2\delta_{\alpha\beta}-p_\alpha p_\beta}{k^2(k^2-p^2+i\epsilon)}e^{-a_\text{ho}^2p^2/2},
\eal
where, as part of the Sommerfeld prescription to make the Green's function causal, we have included an infinitesimal imaginary number with $\epsilon>0$ in the denominator to move the poles off from the real axis \cite{Chew1995}. The six independent components are given by
\bal\label{WeylGreensRegularizedApp}
g^*_{xx} &=& (k^2-p_x^2)\mathcal{ I}_0,\nonumber\\
g^*_{yy} &=& (k^2-p_y^2)\mathcal{ I}_0,\nonumber\\
g^*_{zz} &=& (k^2\mathcal I_0-\mathcal I_2),\nonumber\\
g^*_{xy} &=& g^*_{yx} = -p_xp_y\mathcal{ I}_0\nonumber\\
g^*_{xz} &=& g^*_{zx}= -p_x \mathcal{ I}_1,\nonumber\\
g^*_{yz} &=& g^*_{zy} = -p_y \mathcal{ I}_1
\eal \\
where the integrals are given by
\be
\mathcal{I}_0=\mathcal C \int dp_z\fr{e^{-a_\text{ho}^2p_z^2/2}}{(k^2-p_x^2-p_y^2)+i\epsilon-p_z^2},
\ee
\be
\mathcal{I}_1=\mathcal C \int dp_z\fr{p_ze^{-a_\text{ho}^2p_z^2/2}}{(k^2-p_x^2-p_y^2)+i\epsilon-p_z^2},
\ee
and
\be
\mathcal{I}_2=\mathcal C \int dp_z\fr{p_z^2e^{-a_\text{ho}^2p_z^2/2}}{(k^2-p_x^2-p_y^2)+i\epsilon-p_z^2},
\ee
where 
\be
\mathcal C(p_x,p_y) = \fr{1}{2\pi k^2}e^{-a_\text{ho}^2(p_x^2+p_y^2)/2}.
\ee
Using the closed-form solutions for these integrals (see e.g. \cite{Abramowitz1970}), we obtain Eqs.~\eqref{I0} and \eqref{I2}.

\vspace{8 mm}

%%%%%%%%%%%%%%%%%%%%%%%%%%%%%%%%
\section{Cut-off independence of Green's regularization}\label{cutoffIndependence}
%%%%%%%%%%%%%%%%%%%%%%%%%%%%%%%%
As described in Ref.~\cite{Antezza2009}, in order to demonstrate that \refeq{cutoffIndependent} is independent of $a_\text{ho}$, we need to differentiate \refeq{cutoffIndependent} with respect to $a_\text{ho}^2$ and show that the value of the resulting derivative goes to zero as the $a_\text{ho}\to 0$ limit is taken. To simplify the calculations, we use the form of the Green's function as given in \refeq{ScalarGreensDerivative} and the following observation \cite{Chew1995,Dung1998}
\be
\fr{e^{ikr}}{4\pi r}=-\int\fr{d^3\bp}{(2\pi)^3}\fr{e^{i\bp\cdot\br}}{k^2-p^2}.
\ee
We substitute into \refeq{cutoffIndependent} the following expression
\bal
&&G^*_{0,\alpha\beta}(\omega,\mathbf r)\nonumber\\
&&\qquad=\left[\delta_{\alpha\beta}+\fr{\pa_{r_\alpha}\pa_{ r_\beta}}{k^2}\right]\int\fr{d^3\bp}{(2\pi)^3}\fr{e^{i\bp\cdot\br}}{k^2-p^2}e^{-a_\text{ho}^2p^2/2}.\qquad 
\eal
After taking the derivative with respect to $a_\text{ho}^2$ and performing the resulting Gaussian integral, we obtain
\begin{widetext}
\bal
\lim\limits_{a_\text{ho}\to 0 }\pa_{a_\text{ho}^2}\left[e^{k^2a_\text{ho}^2}\sum\limits_{\mathbf R\neq 0}e^{i\bk_B\cdot \mathbf R}G^*_{\alpha\beta}(\mathbf R)\right]
=\lim\limits_{a_\text{ho}\to 0 }\sum\limits_{\mathbf R\neq 0}e^{i\bk_B\cdot \mathbf R}\left(\delta_{\alpha\beta}+\fr{\pa_{r_\alpha}\pa_{ r_\beta}}{k^2}\right)\fr{e^{a_\text{ho}^2k^2}}{2}\left(\fr{2\pi}{a^2_\text{ho}}\right)^{3/2}e^{-\fr{R^2}{4a_\text{\tiny ho}^2}}=0,
\eal
\end{widetext}
where the last equality follows from the observation that ${\exp(-R^2/4a_\text{\tiny ho}^2})\to 0$ as $a_\text{ho}\to 0$ for $R\neq 0$.

%%%%%%%%%%%%%%%%%%%%%%%%%%%%%%%%%%%%%
\section{Green's function near a flat surface}\label{greensNearFlatSurface}
%%%%%%%%%%%%%%%%%%%%%%%%%%%%%%%%

The Weyl decomposition of the Green's function in the presence of flat surface has previously been derived by a number authors 
\cite{Dung1998,Sipe1987,Tomas1995,Archambault2009}. Assuming the interface is located at $z=-h$, for $z>-h$ the Weyl decomposition  is given by
\bal\label{greensWeylNearPlanarSurface}
g_{\alpha\beta}(\bp;z)=g_{0,\alpha\beta}(\bp;z)+g_{\text{sc},\alpha\beta}^{(1)}(\bp;z),
\eal
where $g_{0,\alpha\beta}(\bp;z)$ is the Green's function in free space discussed previously and the term accounting for the scattering from the surface is given by
\bal \label{SurfaceGreens}
g_{\text{sc},\alpha\beta}^{(1)}(\bp;z)=\frac{-ie^{ik_{d}(2h+z)}}{2k_{d}}\left[r_S\hat {\mbf S }_\alpha \hat {\mbf S }_\beta+r_P\hat {\mbf P }^-_\alpha \hat{\mbf P }^+_\beta\right],\qquad
\eal
where the Fresnel coefficients are given by
\be
r_S=\fr{k_{d}-k_{m}}{k_{d}+k_{m}}\quad \text{and} \quad r_P=\fr{\vare_mk_{d}-\vare_dk_{m}}{\vare_mk_{d}+\vare_dk_{m}},
\ee
with 
\bal
k_{d}=\left(\vare_dk^2-p_x^2-p_y^2\right)^{1/2}\; \geq 0
\eal
and
\bal
k_{m}=\left(\vare_mk^2-p_x^2-p_y^2\right)^{1/2}\;\geq 0,
\eal
where $k=\omega/c$. The unit vectors for the $S$- and ${P\text{-polarizations}}$ are given by:
\bal
\hat {\mbf S}&=&\fr{1}{p}(p_y\hat x-p_x\hat y),\\
\hat {\mbf P}^\pm&=&\fr{1}{k\sqrt{\vare_d}}\left(\sqrt{p_x^2+p_y^2}\hat z\mp k_{d}\fr{p_x\hat x+p_y\hat y}{\sqrt{p_x^2+p_y^2}}\right).
\eal

Using these expressions, we obtain the following matrix expressions

\begin{equation}
\hat {\mbf S } \hat {\mbf S }=
\renewcommand\arraystretch{2}
 \begin{bmatrix}
    \fr{p_y^2}{p_x^2+p_y^2} & -\fr{p_xp_y}{p_x^2+p_y^2} & 0 \\
   -\fr{p_xp_y}{p_x^2+p_y^2} & \fr{p_x^2}{p_x^2+p_y^2} & 0 \\
    0 & 0 & 0
  \end{bmatrix},
\end{equation}
 and 
\begin{equation}
\hat {\mbf P }^- \hat{\mbf P }^+=
\renewcommand\arraystretch{2}
 \begin{bmatrix}
  -\fr{k_{d}^2}{\vare_dk^2}\fr{p_x^2}{p_x^2+p_y^2} & -\fr{k_{d}^2}{\vare_dk^2}\fr{p_xp_y}{p_x^2+p_y^2}  & -\fr{k_{d}}{\vare_dk^2}p_x \\
   -\fr{k_{d}^2}{\vare_dk^2}\fr{p_xp_y}{p_x^2+p_y^2} & -\fr{k_{d}^2}{\vare_dk^2}\fr{p_y^2}{p_x^2+p_y^2}& -\fr{k_{d}}{\vare_dk^2}p_y \\
   \fr{k_{d}}{\vare_dk^2}p_x & \fr{k_{d}}{\vare_dk^2}p_y & \fr{p_x^2+p_y^2}{\vare_dk^2}
  \end{bmatrix}.
\end{equation} 

Therefore, the full matrix expression in Cartesian coordinates for the scattered part of the Green's function is given by

\begin{widetext}
\bal\label{ScatteredGreensMomentum}\nonumber
&&\mbf {g}_\text{sc}(\bp;z)=
%-\frac{i}{2k_{d}}\left[r_S\hat {\mbf S } \hat {\mbf S }+r_P\hat {\mbf P }^- \hat{\mbf P }^+\right]e^{ik_{d}(2d+z)}\\&&=
-\fr{i}{2k_{d}}e^{ik_{d}(2d+z)}
\renewcommand\arraystretch{3}
\setlength{\arraycolsep}{12pt}
 \begin{bmatrix}
  \fr{p_y^2}{p_x^2+p_y^2}\,r_S-\fr{k_{d}^2}{\vare_dk^2}\fr{p_x^2}{p_x^2+p_y^2}\,r_P &   -\fr{p_xp_y}{p_x^2+p_y^2}\,r_S-\fr{k_{d}^2}{\vare_dk^2}\fr{p_xp_y}{p_x^2+p_y^2}\,r_P  & -\fr{k_{d}}{\vare_dk^2}p_x\,r_P \\
-\fr{p_xp_y}{p_x^2+p_y^2}\,r_S-\fr{k_{d}^2}{\vare_dk^2}\fr{p_xp_y}{p_x^2+p_y^2}\,r_P & \fr{p_x^2}{p_x^2+p_y^2}\,r_S-\fr{k_{d}^2}{\vare_dk^2}\fr{p_y^2}{p_x^2+p_y^2}\,r_P& -\fr{k_{d}}{\vare_dk^2}p_y\,r_P \\
   \fr{k_{d}}{\vare_dk^2}p_x\,r_P & \fr{k_{d}}{\vare_dk^2}p_y\,r_P & \fr{p_x^2+p_y^2}{\vare_dk^2}\,r_P
  \end{bmatrix}.
\eal
\end{widetext}

For completeness, we note that for $z<-h$ the Weyl decomposition takes the form
\bal\label{greensWeylNearPlanarSurfaceBelow}
g_{\alpha\beta}(\bp;z)=g_{\text{sc},\alpha\beta}^{(2)}(\bp;z),
\eal
 where $g_{\text{sc},\alpha\beta}^{(2)}$ is given by {\cite{Dung1998,Sipe1987,Tomas1995,Archambault2009}} 
\bal\label{greensWeylBelowPlanarSurface}
&&g_{\text{sc},\alpha\beta}(\bp;z)=\nonumber\\
&&\qquad\qquad\frac{ie^{i(k_{d}h-k_{m}(h+z))}}{2k_{d}}\left[t_S\hat {\mbf S }_\alpha \hat {\mbf S }_\beta+t_P\hat {\mbf P }^{m}_\alpha \hat{\mbf P }^+_\beta\right],\qquad\;\;
\eal
where the Fresnel coefficients are given by
\be
t_S=\fr{2k_{d}}{k_{d}+k_{m}}\quad \text{and} \quad t_P=\fr{2k_{d}\sqrt{\vare_d\vare_m}}{\vare_mk_{d}+\vare_dk_{m}},
\ee
and 
\bal
\hat {\mbf P}^m=\fr{1}{k\sqrt{\vare_m}}\left((p_x^2+p_y^2)\,\hat z- k_{m}\fr{p_x\hat x+p_y\hat y}{\sqrt{p_x^2+p_y^2}}\right).\quad
\eal

Furthermore, we note that our method can be extended to describe atoms in the presence of multiple interfaces in a straightforward manner, since the Green's function in {\it momentum space} can be written down in a closed form in the presence of arbitrarily layered planar media. In particular, the above expressions for the Green's function depend on the particular planar geometry only through the Fresnel coefficients. In order to treat a more complicated planar geometry, we simply need to substitute the relevant Fresnel coefficients into \refeq{SurfaceGreens} (see Ref.~\cite{Sipe1987} for more details). 

In order to perform the summation in momentum space, it is also necessary to find an expression for $G_{\text{sc},\alpha\beta}(\mbf 0)$, which can be expressed in terms of the following integral
\be
G_{\text{sc},\alpha\beta}(\mbf 0)=\int \fr{dp_x\, dp_y}{(2\pi)^2}  g_{\text{sc},\alpha\beta}(\bp;0).
\ee
After performing the angular integral in the $p_x$-$p_y$ plane, only the diagonal terms survive and we are left with the components  
\be
G_\text{sc,xx}(\mbf 0)=-\fr{i}{8\pi}k\left[ \mathcal{I}_{xx,s}-\fr{1}{\vare_d}\mathcal{I}_{xx,p} \right]
\ee 
and
\be
G_\text{sc,yy}(\mbf 0)=G_\text{sc,xx}(\mbf 0),
\ee
and
\be
G_\text{sc,zz}(\mbf 0)=-\fr{i}{4\pi}\fr{1}{\vare_d}k\, \mathcal{I}_{zz,p},
\ee 
where the dimensionless contour integrals are given by
\be
\mathcal{I}_{xx,s}=\int_0^\infty dx\fr{x}{\Lambda_d}\fr{\Lambda_d-\Lambda_m}{\Lambda_d+\Lambda_m}e^{i\Lambda_d2kd},
\ee
and
\be
\mathcal{I}_{xx,p}=\int_0^\infty dx\;x\,\Lambda_d\fr{\vare_m\Lambda_d-\vare_d\Lambda_m}{\vare_m\Lambda_d+\vare_d\Lambda_m}e^{i\Lambda_d2kd},
\ee
and
\be
\mathcal{I}_{zz,p}=\int_0^\infty dx\fr{x^3}{\Lambda_d}\fr{\vare_m\Lambda_d-\vare_d\Lambda_m}{\vare_m\Lambda_d+\vare_d\Lambda_m}e^{i\Lambda_d2kd}.
\ee
The $x$-dependent functions $\Lambda_d$ and $\Lambda_m$ are given by
\be
\Lambda_d(x)=\sqrt{\vare_d-x^2},
\ee
and
\be
\Lambda_m(x)=\sqrt{\vare_m-x^2},
\ee
where the square root with ${\text{Re}(\Lambda_{d})\geq 0}$, ${\text{Im}(\Lambda_{d})\geq 0}$ and $\text{Im}(\Lambda_{m})\geq 0$, $\text{Re}(\Lambda_{m})\geq 0$ is taken to preserve causality.
Note that these integral have previously been obtained in the context of studying qubit relaxation rates near metallic surfaces \cite{Henkel1999,Langsjoen2012,Langsjoen2014}.

%\bibliographystyle{apsrev4-1}
%\bibliography{references}

\begin{thebibliography}{55}%
\makeatletter
\providecommand \@ifxundefined [1]{%
 \@ifx{#1\undefined}
}%
\providecommand \@ifnum [1]{%
 \ifnum #1\expandafter \@firstoftwo
 \else \expandafter \@secondoftwo
 \fi
}%
\providecommand \@ifx [1]{%
 \ifx #1\expandafter \@firstoftwo
 \else \expandafter \@secondoftwo
 \fi
}%
\providecommand \natexlab [1]{#1}%
\providecommand \enquote  [1]{``#1''}%
\providecommand \bibnamefont  [1]{#1}%
\providecommand \bibfnamefont [1]{#1}%
\providecommand \citenamefont [1]{#1}%
\providecommand \href@noop [0]{\@secondoftwo}%
\providecommand \href [0]{\begingroup \@sanitize@url \@href}%
\providecommand \@href[1]{\@@startlink{#1}\@@href}%
\providecommand \@@href[1]{\endgroup#1\@@endlink}%
\providecommand \@sanitize@url [0]{\catcode `\\12\catcode `\$12\catcode
  `\&12\catcode `\#12\catcode `\^12\catcode `\_12\catcode `\%12\relax}%
\providecommand \@@startlink[1]{}%
\providecommand \@@endlink[0]{}%
\providecommand \url  [0]{\begingroup\@sanitize@url \@url }%
\providecommand \@url [1]{\endgroup\@href {#1}{\urlprefix }}%
\providecommand \urlprefix  [0]{URL }%
\providecommand \Eprint [0]{\href }%
\providecommand \doibase [0]{http://dx.doi.org/}%
\providecommand \selectlanguage [0]{\@gobble}%
\providecommand \bibinfo  [0]{\@secondoftwo}%
\providecommand \bibfield  [0]{\@secondoftwo}%
\providecommand \translation [1]{[#1]}%
\providecommand \BibitemOpen [0]{}%
\providecommand \bibitemStop [0]{}%
\providecommand \bibitemNoStop [0]{.\EOS\space}%
\providecommand \EOS [0]{\spacefactor3000\relax}%
\providecommand \BibitemShut  [1]{\csname bibitem#1\endcsname}%
\let\auto@bib@innerbib\@empty
%</preamble>
\bibitem [{\citenamefont {Deutsch}\ \emph {et~al.}(1995)\citenamefont
  {Deutsch}, \citenamefont {Spreeuw}, \citenamefont {Rolston},\ and\
  \citenamefont {Phillips}}]{Deutsch1995}%
  \BibitemOpen
  \bibfield  {author} {\bibinfo {author} {\bibfnamefont {I.~H.}\ \bibnamefont
  {Deutsch}}, \bibinfo {author} {\bibfnamefont {R.~J.~C.}\ \bibnamefont
  {Spreeuw}}, \bibinfo {author} {\bibfnamefont {S.~L.}\ \bibnamefont
  {Rolston}}, \ and\ \bibinfo {author} {\bibfnamefont {W.~D.}\ \bibnamefont
  {Phillips}},\ }\href {https://link.aps.org/doi/10.1103/PhysRevA.52.1394}
  {\bibfield  {journal} {\bibinfo  {journal} {Physical Review A}\ }\textbf
  {\bibinfo {volume} {52}},\ \bibinfo {pages} {1394} (\bibinfo {year}
  {1995})}\BibitemShut {NoStop}%
\bibitem [{\citenamefont {van Coevorden}\ \emph {et~al.}(1996)\citenamefont
  {van Coevorden}, \citenamefont {Sprik}, \citenamefont {Tip},\ and\
  \citenamefont {Lagendijk}}]{VanCoevorden1996}%
  \BibitemOpen
  \bibfield  {author} {\bibinfo {author} {\bibfnamefont {D.}~\bibnamefont {van
  Coevorden}}, \bibinfo {author} {\bibfnamefont {R.}~\bibnamefont {Sprik}},
  \bibinfo {author} {\bibfnamefont {A.}~\bibnamefont {Tip}}, \ and\ \bibinfo
  {author} {\bibfnamefont {A.}~\bibnamefont {Lagendijk}},\ }\href
  {http://journals.aps.org/prl/abstract/10.1103/PhysRevLett.77.2412} {\bibfield
   {journal} {\bibinfo  {journal} {Physical Review Letters}\ }\textbf {\bibinfo
  {volume} {77}},\ \bibinfo {pages} {2412} (\bibinfo {year}
  {1996})}\BibitemShut {NoStop}%
\bibitem [{\citenamefont {de~Vries}\ \emph {et~al.}(1998)\citenamefont
  {de~Vries}, \citenamefont {van Coevorden},\ and\ \citenamefont
  {Lagendijk}}]{DeVries1998}%
  \BibitemOpen
  \bibfield  {author} {\bibinfo {author} {\bibfnamefont {P.}~\bibnamefont
  {de~Vries}}, \bibinfo {author} {\bibfnamefont {D.}~\bibnamefont {van
  Coevorden}}, \ and\ \bibinfo {author} {\bibfnamefont {A.}~\bibnamefont
  {Lagendijk}},\ }\href@noop {} {\bibfield  {journal} {\bibinfo  {journal}
  {Reviews of Modern Physics}\ }\textbf {\bibinfo {volume} {70}},\ \bibinfo
  {pages} {447} (\bibinfo {year} {1998})}\BibitemShut {NoStop}%
\bibitem [{\citenamefont {Klugkist}\ \emph {et~al.}(2006)\citenamefont
  {Klugkist}, \citenamefont {Mostovoy},\ and\ \citenamefont
  {Knoester}}]{Klugkist2006}%
  \BibitemOpen
  \bibfield  {author} {\bibinfo {author} {\bibfnamefont {J.~A.}\ \bibnamefont
  {Klugkist}}, \bibinfo {author} {\bibfnamefont {M.}~\bibnamefont {Mostovoy}},
  \ and\ \bibinfo {author} {\bibfnamefont {J.}~\bibnamefont {Knoester}},\
  }\href {http://link.aps.org/doi/10.1103/PhysRevLett.96.163903} {\bibfield
  {journal} {\bibinfo  {journal} {Physical Review Letters}\ }\textbf {\bibinfo
  {volume} {96}},\ \bibinfo {pages} {163903} (\bibinfo {year}
  {2006})}\BibitemShut {NoStop}%
\bibitem [{\citenamefont {Antezza}\ and\ \citenamefont
  {Castin}(2009{\natexlab{a}})}]{Antezza2009}%
  \BibitemOpen
  \bibfield  {author} {\bibinfo {author} {\bibfnamefont {M.}~\bibnamefont
  {Antezza}}\ and\ \bibinfo {author} {\bibfnamefont {Y.}~\bibnamefont
  {Castin}},\ }\href {http://link.aps.org/doi/10.1103/PhysRevLett.103.123903}
  {\bibfield  {journal} {\bibinfo  {journal} {Physical Review Letters}\
  }\textbf {\bibinfo {volume} {103}},\ \bibinfo {pages} {123903} (\bibinfo
  {year} {2009}{\natexlab{a}})}\BibitemShut {NoStop}%
\bibitem [{\citenamefont {Antezza}\ and\ \citenamefont
  {Castin}(2009{\natexlab{b}})}]{Antezza2009a}%
  \BibitemOpen
  \bibfield  {author} {\bibinfo {author} {\bibfnamefont {M.}~\bibnamefont
  {Antezza}}\ and\ \bibinfo {author} {\bibfnamefont {Y.}~\bibnamefont
  {Castin}},\ }\href {http://link.aps.org/doi/10.1103/PhysRevA.80.013816}
  {\bibfield  {journal} {\bibinfo  {journal} {Physical Review A}\ }\textbf
  {\bibinfo {volume} {80}},\ \bibinfo {pages} {013816} (\bibinfo {year}
  {2009}{\natexlab{b}})}\BibitemShut {NoStop}%
\bibitem [{\citenamefont {Bienaim{\'{e}}}\ \emph {et~al.}(2012)\citenamefont
  {Bienaim{\'{e}}}, \citenamefont {Piovella},\ and\ \citenamefont
  {Kaiser}}]{Bienaime2012}%
  \BibitemOpen
  \bibfield  {author} {\bibinfo {author} {\bibfnamefont {T.}~\bibnamefont
  {Bienaim{\'{e}}}}, \bibinfo {author} {\bibfnamefont {N.}~\bibnamefont
  {Piovella}}, \ and\ \bibinfo {author} {\bibfnamefont {R.}~\bibnamefont
  {Kaiser}},\ }\href {http://link.aps.org/doi/10.1103/PhysRevLett.108.123602}
  {\bibfield  {journal} {\bibinfo  {journal} {Physical Review Letters}\
  }\textbf {\bibinfo {volume} {108}},\ \bibinfo {pages} {123602} (\bibinfo
  {year} {2012})}\BibitemShut {NoStop}%
\bibitem [{\citenamefont {Guerin}\ \emph {et~al.}(2016)\citenamefont {Guerin},
  \citenamefont {Ara{\'{u}}jo},\ and\ \citenamefont {Kaiser}}]{Guerin2016}%
  \BibitemOpen
  \bibfield  {author} {\bibinfo {author} {\bibfnamefont {W.}~\bibnamefont
  {Guerin}}, \bibinfo {author} {\bibfnamefont {M.~O.}\ \bibnamefont
  {Ara{\'{u}}jo}}, \ and\ \bibinfo {author} {\bibfnamefont {R.}~\bibnamefont
  {Kaiser}},\ }\href {http://link.aps.org/doi/10.1103/PhysRevLett.116.083601}
  {\bibfield  {journal} {\bibinfo  {journal} {Physical Review Letters}\
  }\textbf {\bibinfo {volume} {116}},\ \bibinfo {pages} {083601} (\bibinfo
  {year} {2016})}\BibitemShut {NoStop}%
\bibitem [{\citenamefont {Perczel}\ \emph {et~al.}(2017)\citenamefont
  {Perczel}, \citenamefont {Borregaard}, \citenamefont {Chang}, \citenamefont
  {Pichler}, \citenamefont {Yelin}, \citenamefont {Zoller},\ and\ \citenamefont
  {Lukin}}]{Perczel2017}%
  \BibitemOpen
  \bibfield  {author} {\bibinfo {author} {\bibfnamefont {J.}~\bibnamefont
  {Perczel}}, \bibinfo {author} {\bibfnamefont {J.}~\bibnamefont {Borregaard}},
  \bibinfo {author} {\bibfnamefont {D.~E.}\ \bibnamefont {Chang}}, \bibinfo
  {author} {\bibfnamefont {H.}~\bibnamefont {Pichler}}, \bibinfo {author}
  {\bibfnamefont {S.~F.}\ \bibnamefont {Yelin}}, \bibinfo {author}
  {\bibfnamefont {P.}~\bibnamefont {Zoller}}, \ and\ \bibinfo {author}
  {\bibfnamefont {M.~D.}\ \bibnamefont {Lukin}},\ }\href {\doibase
  10.1103/PhysRevLett.119.023603} {\bibfield  {journal} {\bibinfo  {journal}
  {Physical Review Letters}\ }\textbf {\bibinfo {volume} {119}},\ \bibinfo
  {pages} {023603} (\bibinfo {year} {2017})}\BibitemShut {NoStop}%
\bibitem [{\citenamefont {Bettles}\ \emph {et~al.}(2017)\citenamefont
  {Bettles}, \citenamefont {Min{\'{a}}{\v r}}, \citenamefont {Lesanovsky},
  \citenamefont {Adams},\ and\ \citenamefont {Olmos}}]{Bettles2017}%
  \BibitemOpen
  \bibfield  {author} {\bibinfo {author} {\bibfnamefont {R.~J.}\ \bibnamefont
  {Bettles}}, \bibinfo {author} {\bibfnamefont {J.}~\bibnamefont {Min{\'{a}}{\v
  r}}}, \bibinfo {author} {\bibfnamefont {I.}~\bibnamefont {Lesanovsky}},
  \bibinfo {author} {\bibfnamefont {C.~S.}\ \bibnamefont {Adams}}, \ and\
  \bibinfo {author} {\bibfnamefont {B.}~\bibnamefont {Olmos}},\ }\href
  {http://arxiv.org/abs/1703.03351} {\  (\bibinfo {year} {2017})},\ \Eprint
  {http://arxiv.org/abs/1703.03351} {arXiv:1703.03351} \BibitemShut {NoStop}%
\bibitem [{\citenamefont {Bettles}\ \emph {et~al.}(2016)\citenamefont
  {Bettles}, \citenamefont {Gardiner},\ and\ \citenamefont
  {Adams}}]{Bettles2016}%
  \BibitemOpen
  \bibfield  {author} {\bibinfo {author} {\bibfnamefont {R.~J.}\ \bibnamefont
  {Bettles}}, \bibinfo {author} {\bibfnamefont {S.~A.}\ \bibnamefont
  {Gardiner}}, \ and\ \bibinfo {author} {\bibfnamefont {C.~S.}\ \bibnamefont
  {Adams}},\ }\href {http://link.aps.org/doi/10.1103/PhysRevLett.116.103602}
  {\bibfield  {journal} {\bibinfo  {journal} {Physical Review Letters}\
  }\textbf {\bibinfo {volume} {116}},\ \bibinfo {pages} {103602} (\bibinfo
  {year} {2016})}\BibitemShut {NoStop}%
\bibitem [{\citenamefont {Syzranov}\ \emph {et~al.}(2016)\citenamefont
  {Syzranov}, \citenamefont {Wall}, \citenamefont {Zhu}, \citenamefont
  {Gurarie},\ and\ \citenamefont {Rey}}]{Syzranov2016}%
  \BibitemOpen
  \bibfield  {author} {\bibinfo {author} {\bibfnamefont {S.~V.}\ \bibnamefont
  {Syzranov}}, \bibinfo {author} {\bibfnamefont {M.~L.}\ \bibnamefont {Wall}},
  \bibinfo {author} {\bibfnamefont {B.}~\bibnamefont {Zhu}}, \bibinfo {author}
  {\bibfnamefont {V.}~\bibnamefont {Gurarie}}, \ and\ \bibinfo {author}
  {\bibfnamefont {A.~M.}\ \bibnamefont {Rey}},\ }\href
  {http://www.nature.com/doifinder/10.1038/ncomms13543} {\bibfield  {journal}
  {\bibinfo  {journal} {Nature Communications}\ }\textbf {\bibinfo {volume}
  {7}},\ \bibinfo {pages} {13543} (\bibinfo {year} {2016})}\BibitemShut
  {NoStop}%
\bibitem [{\citenamefont {Shahmoon}\ \emph {et~al.}(2017)\citenamefont
  {Shahmoon}, \citenamefont {Wild}, \citenamefont {Lukin},\ and\ \citenamefont
  {Yelin}}]{Shahmoon2017}%
  \BibitemOpen
  \bibfield  {author} {\bibinfo {author} {\bibfnamefont {E.}~\bibnamefont
  {Shahmoon}}, \bibinfo {author} {\bibfnamefont {D.~S.}\ \bibnamefont {Wild}},
  \bibinfo {author} {\bibfnamefont {M.~D.}\ \bibnamefont {Lukin}}, \ and\
  \bibinfo {author} {\bibfnamefont {S.~F.}\ \bibnamefont {Yelin}},\ }\href
  {https://journals.aps.org/prl/abstract/10.1103/PhysRevLett.118.113601}
  {\bibfield  {journal} {\bibinfo  {journal} {Physical Review Letters}\
  }\textbf {\bibinfo {volume} {118}},\ \bibinfo {pages} {113601} (\bibinfo
  {year} {2017})}\BibitemShut {NoStop}%
\bibitem [{\citenamefont {Asenjo-Garcia}\ \emph {et~al.}(2017)\citenamefont
  {Asenjo-Garcia}, \citenamefont {Moreno-Cardoner}, \citenamefont {Albrecht},
  \citenamefont {Kimble},\ and\ \citenamefont {Chang}}]{Asenjo-Garcia2017}%
  \BibitemOpen
  \bibfield  {author} {\bibinfo {author} {\bibfnamefont {A.}~\bibnamefont
  {Asenjo-Garcia}}, \bibinfo {author} {\bibfnamefont {M.}~\bibnamefont
  {Moreno-Cardoner}}, \bibinfo {author} {\bibfnamefont {A.}~\bibnamefont
  {Albrecht}}, \bibinfo {author} {\bibfnamefont {H.~J.}\ \bibnamefont
  {Kimble}}, \ and\ \bibinfo {author} {\bibfnamefont {D.~E.}\ \bibnamefont
  {Chang}},\ }\href {\doibase 10.1103/PhysRevX.7.031024} {\bibfield  {journal}
  {\bibinfo  {journal} {Physical Review X}\ }\textbf {\bibinfo {volume} {7}},\
  \bibinfo {pages} {031024} (\bibinfo {year} {2017})}\BibitemShut {NoStop}%
\bibitem [{\citenamefont {Facchinetti}\ \emph {et~al.}(2016)\citenamefont
  {Facchinetti}, \citenamefont {Jenkins},\ and\ \citenamefont
  {Ruostekoski}}]{Facchinetti2016}%
  \BibitemOpen
  \bibfield  {author} {\bibinfo {author} {\bibfnamefont {G.}~\bibnamefont
  {Facchinetti}}, \bibinfo {author} {\bibfnamefont {S.~D.}\ \bibnamefont
  {Jenkins}}, \ and\ \bibinfo {author} {\bibfnamefont {J.}~\bibnamefont
  {Ruostekoski}},\ }\href
  {https://link.aps.org/doi/10.1103/PhysRevLett.117.243601} {\bibfield
  {journal} {\bibinfo  {journal} {Physical Review Letters}\ }\textbf {\bibinfo
  {volume} {117}},\ \bibinfo {pages} {243601} (\bibinfo {year}
  {2016})}\BibitemShut {NoStop}%
\bibitem [{\citenamefont {Collin}(2004)}]{Collin2004}%
  \BibitemOpen
  \bibfield  {author} {\bibinfo {author} {\bibfnamefont {R.}~\bibnamefont
  {Collin}},\ }\href {http://ieeexplore.ieee.org/document/1305535/} {\bibfield
  {journal} {\bibinfo  {journal} {IEEE Antennas and Propagation Magazine}\
  }\textbf {\bibinfo {volume} {46}},\ \bibinfo {pages} {64} (\bibinfo {year}
  {2004})}\BibitemShut {NoStop}%
\bibitem [{\citenamefont {Ebbesen}\ \emph {et~al.}(1998)\citenamefont
  {Ebbesen}, \citenamefont {Lezec}, \citenamefont {Ghaemi}, \citenamefont
  {Thio},\ and\ \citenamefont {Wolff}}]{Ebbesen1998}%
  \BibitemOpen
  \bibfield  {author} {\bibinfo {author} {\bibfnamefont {T.~W.}\ \bibnamefont
  {Ebbesen}}, \bibinfo {author} {\bibfnamefont {H.~J.}\ \bibnamefont {Lezec}},
  \bibinfo {author} {\bibfnamefont {H.~F.}\ \bibnamefont {Ghaemi}}, \bibinfo
  {author} {\bibfnamefont {T.}~\bibnamefont {Thio}}, \ and\ \bibinfo {author}
  {\bibfnamefont {P.~A.}\ \bibnamefont {Wolff}},\ }\href
  {http://www.nature.com/doifinder/10.1038/35570} {\bibfield  {journal}
  {\bibinfo  {journal} {Nature}\ }\textbf {\bibinfo {volume} {391}},\ \bibinfo
  {pages} {667} (\bibinfo {year} {1998})}\BibitemShut {NoStop}%
\bibitem [{\citenamefont {Chang}\ \emph {et~al.}(2006)\citenamefont {Chang},
  \citenamefont {S{\o}rensen}, \citenamefont {Hemmer},\ and\ \citenamefont
  {Lukin}}]{Chang2006}%
  \BibitemOpen
  \bibfield  {author} {\bibinfo {author} {\bibfnamefont {D.~E.}\ \bibnamefont
  {Chang}}, \bibinfo {author} {\bibfnamefont {A.~S.}\ \bibnamefont
  {S{\o}rensen}}, \bibinfo {author} {\bibfnamefont {P.~R.}\ \bibnamefont
  {Hemmer}}, \ and\ \bibinfo {author} {\bibfnamefont {M.~D.}\ \bibnamefont
  {Lukin}},\ }\href {http://link.aps.org/doi/10.1103/PhysRevLett.97.053002}
  {\bibfield  {journal} {\bibinfo  {journal} {Physical Review Letters}\
  }\textbf {\bibinfo {volume} {97}},\ \bibinfo {pages} {053002} (\bibinfo
  {year} {2006})}\BibitemShut {NoStop}%
\bibitem [{\citenamefont {{Garc{\'{i}}a de Abajo}}(2007)}]{GarciadeAbajo2007}%
  \BibitemOpen
  \bibfield  {author} {\bibinfo {author} {\bibfnamefont {F.~J.}\ \bibnamefont
  {{Garc{\'{i}}a de Abajo}}},\ }\href
  {http://link.aps.org/doi/10.1103/RevModPhys.79.1267} {\bibfield  {journal}
  {\bibinfo  {journal} {Reviews of Modern Physics}\ }\textbf {\bibinfo {volume}
  {79}},\ \bibinfo {pages} {1267} (\bibinfo {year} {2007})}\BibitemShut
  {NoStop}%
\bibitem [{\citenamefont {Nikitin}\ \emph {et~al.}(2010)\citenamefont
  {Nikitin}, \citenamefont {Garc{\'{i}}a-Vidal},\ and\ \citenamefont
  {Mart{\'{i}}n-Moreno}}]{Nikitin2010}%
  \BibitemOpen
  \bibfield  {author} {\bibinfo {author} {\bibfnamefont {A.~Y.}\ \bibnamefont
  {Nikitin}}, \bibinfo {author} {\bibfnamefont {F.~J.}\ \bibnamefont
  {Garc{\'{i}}a-Vidal}}, \ and\ \bibinfo {author} {\bibfnamefont
  {L.}~\bibnamefont {Mart{\'{i}}n-Moreno}},\ }\href
  {https://link.aps.org/doi/10.1103/PhysRevLett.105.073902} {\bibfield
  {journal} {\bibinfo  {journal} {Physical Review Letters}\ }\textbf {\bibinfo
  {volume} {105}},\ \bibinfo {pages} {073902} (\bibinfo {year}
  {2010})}\BibitemShut {NoStop}%
\bibitem [{\citenamefont {Kildishev}\ \emph {et~al.}(2013)\citenamefont
  {Kildishev}, \citenamefont {Boltasseva},\ and\ \citenamefont
  {Shalaev}}]{Kildishev2013}%
  \BibitemOpen
  \bibfield  {author} {\bibinfo {author} {\bibfnamefont {A.~V.}\ \bibnamefont
  {Kildishev}}, \bibinfo {author} {\bibfnamefont {A.}~\bibnamefont
  {Boltasseva}}, \ and\ \bibinfo {author} {\bibfnamefont {V.~M.}\ \bibnamefont
  {Shalaev}},\ }\href@noop {} {\bibfield  {journal} {\bibinfo  {journal}
  {Science}\ }\textbf {\bibinfo {volume} {339}} (\bibinfo {year}
  {2013})}\BibitemShut {NoStop}%
\bibitem [{\citenamefont {Tame}\ \emph {et~al.}(2013)\citenamefont {Tame},
  \citenamefont {Mcenery}, \citenamefont {{\"{O}}zdemir}, \citenamefont {Lee},
  \citenamefont {Maier},\ and\ \citenamefont {Kim}}]{Tame2013}%
  \BibitemOpen
  \bibfield  {author} {\bibinfo {author} {\bibfnamefont {M.~S.}\ \bibnamefont
  {Tame}}, \bibinfo {author} {\bibfnamefont {K.~R.}\ \bibnamefont {Mcenery}},
  \bibinfo {author} {\bibfnamefont {{\c S}.~K.}\ \bibnamefont {{\"{O}}zdemir}},
  \bibinfo {author} {\bibfnamefont {J.}~\bibnamefont {Lee}}, \bibinfo {author}
  {\bibfnamefont {S.~A.}\ \bibnamefont {Maier}}, \ and\ \bibinfo {author}
  {\bibfnamefont {M.~S.}\ \bibnamefont {Kim}},\ }\href
  {http://dx.doi.org/10.1038/nphys2615} {\bibfield  {journal} {\bibinfo
  {journal} {Nature Physics}\ }\textbf {\bibinfo {volume} {9}},\ \bibinfo
  {pages} {329} (\bibinfo {year} {2013})}\BibitemShut {NoStop}%
\bibitem [{\citenamefont {Delga}\ \emph {et~al.}(2014)\citenamefont {Delga},
  \citenamefont {Feist}, \citenamefont {Bravo-Abad},\ and\ \citenamefont
  {Garcia-Vidal}}]{Delga2014}%
  \BibitemOpen
  \bibfield  {author} {\bibinfo {author} {\bibfnamefont {A.}~\bibnamefont
  {Delga}}, \bibinfo {author} {\bibfnamefont {J.}~\bibnamefont {Feist}},
  \bibinfo {author} {\bibfnamefont {J.}~\bibnamefont {Bravo-Abad}}, \ and\
  \bibinfo {author} {\bibfnamefont {F.}~\bibnamefont {Garcia-Vidal}},\ }\href
  {http://link.aps.org/doi/10.1103/PhysRevLett.112.253601} {\bibfield
  {journal} {\bibinfo  {journal} {Physical Review Letters}\ }\textbf {\bibinfo
  {volume} {112}},\ \bibinfo {pages} {253601} (\bibinfo {year}
  {2014})}\BibitemShut {NoStop}%
\bibitem [{\citenamefont {High}\ \emph {et~al.}(2015)\citenamefont {High},
  \citenamefont {Devlin}, \citenamefont {Dibos}, \citenamefont {Polking},
  \citenamefont {Wild}, \citenamefont {Perczel}, \citenamefont {de~Leon},
  \citenamefont {Lukin},\ and\ \citenamefont {Park}}]{High2015}%
  \BibitemOpen
  \bibfield  {author} {\bibinfo {author} {\bibfnamefont {A.~A.}\ \bibnamefont
  {High}}, \bibinfo {author} {\bibfnamefont {R.~C.}\ \bibnamefont {Devlin}},
  \bibinfo {author} {\bibfnamefont {A.}~\bibnamefont {Dibos}}, \bibinfo
  {author} {\bibfnamefont {M.}~\bibnamefont {Polking}}, \bibinfo {author}
  {\bibfnamefont {D.~S.}\ \bibnamefont {Wild}}, \bibinfo {author}
  {\bibfnamefont {J.}~\bibnamefont {Perczel}}, \bibinfo {author} {\bibfnamefont
  {N.~P.}\ \bibnamefont {de~Leon}}, \bibinfo {author} {\bibfnamefont {M.~D.}\
  \bibnamefont {Lukin}}, \ and\ \bibinfo {author} {\bibfnamefont
  {H.}~\bibnamefont {Park}},\ }\href
  {http://www.nature.com/doifinder/10.1038/nature14477} {\bibfield  {journal}
  {\bibinfo  {journal} {Nature}\ }\textbf {\bibinfo {volume} {522}},\ \bibinfo
  {pages} {192} (\bibinfo {year} {2015})}\BibitemShut {NoStop}%
\bibitem [{\citenamefont {Pichler}\ \emph {et~al.}(2015)\citenamefont
  {Pichler}, \citenamefont {Ramos}, \citenamefont {Daley},\ and\ \citenamefont
  {Zoller}}]{Pichler2015}%
  \BibitemOpen
  \bibfield  {author} {\bibinfo {author} {\bibfnamefont {H.}~\bibnamefont
  {Pichler}}, \bibinfo {author} {\bibfnamefont {T.}~\bibnamefont {Ramos}},
  \bibinfo {author} {\bibfnamefont {A.~J.}\ \bibnamefont {Daley}}, \ and\
  \bibinfo {author} {\bibfnamefont {P.}~\bibnamefont {Zoller}},\ }\href
  {http://link.aps.org/doi/10.1103/PhysRevA.91.042116} {\bibfield  {journal}
  {\bibinfo  {journal} {Physical Review A}\ }\textbf {\bibinfo {volume} {91}},\
  \bibinfo {pages} {042116} (\bibinfo {year} {2015})}\BibitemShut {NoStop}%
\bibitem [{\citenamefont {Gardiner}\ and\ \citenamefont
  {Zoller}(2010)}]{Gardiner2010}%
  \BibitemOpen
  \bibfield  {author} {\bibinfo {author} {\bibfnamefont {C.~W.}\ \bibnamefont
  {Gardiner}}\ and\ \bibinfo {author} {\bibfnamefont {P.}~\bibnamefont
  {Zoller}},\ }\href@noop {} {\emph {\bibinfo {title} {{Quantum noise: a
  handbook of Markovian and non-Markovian quantum stochastic methods with
  applications to quantum optics}}}}\ (\bibinfo  {publisher} {Springer},\
  \bibinfo {year} {2010})\BibitemShut {NoStop}%
\bibitem [{\citenamefont {Dalibard}\ \emph {et~al.}(1992)\citenamefont
  {Dalibard}, \citenamefont {Castin},\ and\ \citenamefont
  {M{\o}lmer}}]{Dalibard1992}%
  \BibitemOpen
  \bibfield  {author} {\bibinfo {author} {\bibfnamefont {J.}~\bibnamefont
  {Dalibard}}, \bibinfo {author} {\bibfnamefont {Y.}~\bibnamefont {Castin}}, \
  and\ \bibinfo {author} {\bibfnamefont {K.}~\bibnamefont {M{\o}lmer}},\ }\href
  {http://link.aps.org/doi/10.1103/PhysRevLett.68.580} {\bibfield  {journal}
  {\bibinfo  {journal} {Physical Review Letters}\ }\textbf {\bibinfo {volume}
  {68}},\ \bibinfo {pages} {580} (\bibinfo {year} {1992})}\BibitemShut
  {NoStop}%
\bibitem [{\citenamefont {Carmichael}(1993)}]{Carmichael1993}%
  \BibitemOpen
  \bibfield  {author} {\bibinfo {author} {\bibfnamefont {H.}~\bibnamefont
  {Carmichael}},\ }\href
  {https://books.google.com/books/about/An{\_}Open{\_}Systems{\_}Approach{\_}to{\_}Quantum{\_}Opti.html?id=El5gxgxWhpgC}
  {\emph {\bibinfo {title} {{An open systems approach to quantum optics:
  lectures presented at the Universit{\'e} libre de Bruxelles, October 28 to
  November 4, 1991}}}}\ (\bibinfo  {publisher} {Springer-Verlag},\ \bibinfo
  {year} {1993})\BibitemShut {NoStop}%
\bibitem [{\citenamefont {M{\o}lmer}\ \emph {et~al.}(1993)\citenamefont
  {M{\o}lmer}, \citenamefont {Castin},\ and\ \citenamefont
  {Dalibard}}]{Molmer1993}%
  \BibitemOpen
  \bibfield  {author} {\bibinfo {author} {\bibfnamefont {K.}~\bibnamefont
  {M{\o}lmer}}, \bibinfo {author} {\bibfnamefont {Y.}~\bibnamefont {Castin}}, \
  and\ \bibinfo {author} {\bibfnamefont {J.}~\bibnamefont {Dalibard}},\ }\href
  {https://www.osapublishing.org/abstract.cfm?URI=josab-10-3-524} {\bibfield
  {journal} {\bibinfo  {journal} {Journal of the Optical Society of America B}\
  }\textbf {\bibinfo {volume} {10}},\ \bibinfo {pages} {524} (\bibinfo {year}
  {1993})}\BibitemShut {NoStop}%
\bibitem [{\citenamefont {Bena}\ and\ \citenamefont
  {Montambaux}(2009)}]{Bena2009}%
  \BibitemOpen
  \bibfield  {author} {\bibinfo {author} {\bibfnamefont {C.}~\bibnamefont
  {Bena}}\ and\ \bibinfo {author} {\bibfnamefont {G.}~\bibnamefont
  {Montambaux}},\ }\href
  {http://stacks.iop.org/1367-2630/11/i=9/a=095003?key=crossref.9bfc13536183dcccd2af45d7976ca319}
  {\bibfield  {journal} {\bibinfo  {journal} {New Journal of Physics}\ }\textbf
  {\bibinfo {volume} {11}},\ \bibinfo {pages} {095003} (\bibinfo {year}
  {2009})}\BibitemShut {NoStop}%
\bibitem [{\citenamefont {Dung}\ \emph {et~al.}(1998)\citenamefont {Dung},
  \citenamefont {Kn{\"{o}}ll},\ and\ \citenamefont {Welsch}}]{Dung1998}%
  \BibitemOpen
  \bibfield  {author} {\bibinfo {author} {\bibfnamefont {H.~T.}\ \bibnamefont
  {Dung}}, \bibinfo {author} {\bibfnamefont {L.}~\bibnamefont {Kn{\"{o}}ll}}, \
  and\ \bibinfo {author} {\bibfnamefont {D.-G.}\ \bibnamefont {Welsch}},\
  }\href {http://link.aps.org/doi/10.1103/PhysRevA.57.3931} {\bibfield
  {journal} {\bibinfo  {journal} {Physical Review A}\ }\textbf {\bibinfo
  {volume} {57}},\ \bibinfo {pages} {3931} (\bibinfo {year}
  {1998})}\BibitemShut {NoStop}%
\bibitem [{\citenamefont {Morice}\ \emph {et~al.}(1995)\citenamefont {Morice},
  \citenamefont {Castin},\ and\ \citenamefont {Dalibard}}]{Morice1995}%
  \BibitemOpen
  \bibfield  {author} {\bibinfo {author} {\bibfnamefont {O.}~\bibnamefont
  {Morice}}, \bibinfo {author} {\bibfnamefont {Y.}~\bibnamefont {Castin}}, \
  and\ \bibinfo {author} {\bibfnamefont {J.}~\bibnamefont {Dalibard}},\ }\href
  {http://link.aps.org/doi/10.1103/PhysRevA.51.3896} {\bibfield  {journal}
  {\bibinfo  {journal} {Physical Review A}\ }\textbf {\bibinfo {volume} {51}},\
  \bibinfo {pages} {3896} (\bibinfo {year} {1995})}\BibitemShut {NoStop}%
\bibitem [{\citenamefont {Gross}\ and\ \citenamefont
  {Haroche}(1982)}]{Gross1982}%
  \BibitemOpen
  \bibfield  {author} {\bibinfo {author} {\bibfnamefont {M.}~\bibnamefont
  {Gross}}\ and\ \bibinfo {author} {\bibfnamefont {S.}~\bibnamefont
  {Haroche}},\ }\href
  {http://linkinghub.elsevier.com/retrieve/pii/0370157382901028} {\bibfield
  {journal} {\bibinfo  {journal} {Physics Reports}\ }\textbf {\bibinfo {volume}
  {93}},\ \bibinfo {pages} {301} (\bibinfo {year} {1982})}\BibitemShut
  {NoStop}%
\bibitem [{\citenamefont {Milonni}\ and\ \citenamefont
  {Knight}(1974)}]{Milonni1974}%
  \BibitemOpen
  \bibfield  {author} {\bibinfo {author} {\bibfnamefont {P.~W.}\ \bibnamefont
  {Milonni}}\ and\ \bibinfo {author} {\bibfnamefont {P.~L.}\ \bibnamefont
  {Knight}},\ }\href {https://link.aps.org/doi/10.1103/PhysRevA.10.1096}
  {\bibfield  {journal} {\bibinfo  {journal} {Physical Review A}\ }\textbf
  {\bibinfo {volume} {10}},\ \bibinfo {pages} {1096} (\bibinfo {year}
  {1974})}\BibitemShut {NoStop}%
\bibitem [{\citenamefont {Ashcroft}\ and\ \citenamefont
  {Mermin}(1976)}]{Ashcroft1976}%
  \BibitemOpen
  \bibfield  {author} {\bibinfo {author} {\bibfnamefont {N.~W.}\ \bibnamefont
  {Ashcroft}}\ and\ \bibinfo {author} {\bibfnamefont {N.~D.}\ \bibnamefont
  {Mermin}},\ }\href@noop {} {\emph {\bibinfo {title} {{Solid state
  physics}}}}\ (\bibinfo  {publisher} {Holt, Rinehart and Winston},\ \bibinfo
  {year} {1976})\BibitemShut {NoStop}%
\bibitem [{\citenamefont {Chew}(1999)}]{Chew1995}%
  \BibitemOpen
  \bibfield  {author} {\bibinfo {author} {\bibfnamefont {W.~C.}\ \bibnamefont
  {Chew}},\ }\href
  {http://ieeexplore.ieee.org/xpl/bkabstractplus.jsp?bkn=5270998} {\emph
  {\bibinfo {title} {Wave and fields in inhomogeneous media}}}\ (\bibinfo
  {publisher} {IEEE},\ \bibinfo {year} {1999})\BibitemShut {NoStop}%
\bibitem [{\citenamefont {Joannopoulos}\ \emph {et~al.}(2008)\citenamefont
  {Joannopoulos}, \citenamefont {Johnson}, \citenamefont {Winn},\ and\
  \citenamefont {Meade}}]{Joannopoulos2008}%
  \BibitemOpen
  \bibfield  {author} {\bibinfo {author} {\bibfnamefont {J.~D.}\ \bibnamefont
  {Joannopoulos}}, \bibinfo {author} {\bibfnamefont {S.~G.}\ \bibnamefont
  {Johnson}}, \bibinfo {author} {\bibfnamefont {J.~N.}\ \bibnamefont {Winn}}, \
  and\ \bibinfo {author} {\bibfnamefont {R.~D.}\ \bibnamefont {Meade}},\
  }\href@noop {} {\emph {\bibinfo {title} {{Photonic Crystals: Molding the Flow
  of Light (Second Edition).}}}}\ (\bibinfo  {publisher} {Princeton University
  Press},\ \bibinfo {year} {2008})\BibitemShut {NoStop}%
\bibitem [{\citenamefont {Bernevig}\ and\ \citenamefont
  {Hughes}(2013)}]{Bernevig2013}%
  \BibitemOpen
  \bibfield  {author} {\bibinfo {author} {\bibfnamefont {B.~A.}\ \bibnamefont
  {Bernevig}}\ and\ \bibinfo {author} {\bibfnamefont {T.~L.}\ \bibnamefont
  {Hughes}},\ }\href@noop {} {\emph {\bibinfo {title} {{Topological Insulators
  and Topological Superconductors}}}}\ (\bibinfo  {publisher} {Princeton
  University Press.},\ \bibinfo {year} {2013})\BibitemShut {NoStop}%
\bibitem [{\citenamefont {Fukui}\ \emph {et~al.}(2005)\citenamefont {Fukui},
  \citenamefont {Hatsugai},\ and\ \citenamefont {Suzuki}}]{Fukui2005}%
  \BibitemOpen
  \bibfield  {author} {\bibinfo {author} {\bibfnamefont {T.}~\bibnamefont
  {Fukui}}, \bibinfo {author} {\bibfnamefont {Y.}~\bibnamefont {Hatsugai}}, \
  and\ \bibinfo {author} {\bibfnamefont {H.}~\bibnamefont {Suzuki}},\ }\href
  {http://journals.jps.jp/doi/10.1143/JPSJ.74.1674} {\bibfield  {journal}
  {\bibinfo  {journal} {Journal of the Physical Society of Japan}\ }\textbf
  {\bibinfo {volume} {74}},\ \bibinfo {pages} {1674} (\bibinfo {year}
  {2005})}\BibitemShut {NoStop}%
\bibitem [{Note1()}]{Note1}%
  \BibitemOpen
  \bibinfo {note} {Since the interactions are long range $(\sim 1/r)$, assuming
  periodic boundary conditions along one direction requires the truncation of
  the range of interaction. This leads to small deviations in the decay rates,
  potentially making them negative. This is an artifact of our numerical method
  with no physical significance. When the edge modes are obtained for a lattice
  of atoms that is finite in all directions (see Fig.~\ref {NB_square_time}),
  no such truncation is necessary and all decay rates are
  non-negative.}\BibitemShut {Stop}%
\bibitem [{\citenamefont {Bloch}(2005)}]{Bloch2005}%
  \BibitemOpen
  \bibfield  {author} {\bibinfo {author} {\bibfnamefont {I.}~\bibnamefont
  {Bloch}},\ }\href {http://www.nature.com/doifinder/10.1038/nphys138}
  {\bibfield  {journal} {\bibinfo  {journal} {Nature Physics}\ }\textbf
  {\bibinfo {volume} {1}},\ \bibinfo {pages} {23} (\bibinfo {year}
  {2005})}\BibitemShut {NoStop}%
\bibitem [{\citenamefont {Karzig}\ \emph {et~al.}(2015)\citenamefont {Karzig},
  \citenamefont {Bardyn}, \citenamefont {Lindner},\ and\ \citenamefont
  {Refael}}]{Karzig2015}%
  \BibitemOpen
  \bibfield  {author} {\bibinfo {author} {\bibfnamefont {T.}~\bibnamefont
  {Karzig}}, \bibinfo {author} {\bibfnamefont {C.-E.}\ \bibnamefont {Bardyn}},
  \bibinfo {author} {\bibfnamefont {N.~H.}\ \bibnamefont {Lindner}}, \ and\
  \bibinfo {author} {\bibfnamefont {G.}~\bibnamefont {Refael}},\ }\href
  {http://link.aps.org/doi/10.1103/PhysRevX.5.031001} {\bibfield  {journal}
  {\bibinfo  {journal} {Physical Review X}\ }\textbf {\bibinfo {volume} {5}},\
  \bibinfo {pages} {031001} (\bibinfo {year} {2015})}\BibitemShut {NoStop}%
\bibitem [{Note2()}]{Note2}%
  \BibitemOpen
  \bibinfo {note} {We note that using the exact Green's function in momentum
  space has the other advantage of allowing us to account for the interaction
  of the atoms not only with surface plasmons, but also with other types of
  surface waves, which are collectively known as ``creeping waves'' or
  ``quasi-cylindrical waves'' \cite
  {Lalanne2006,Liu2008,Lalanne2009,Nikitin2010}}\BibitemShut {NoStop}%
\bibitem [{\citenamefont {Frahm}(1983)}]{Frahm1983}%
  \BibitemOpen
  \bibfield  {author} {\bibinfo {author} {\bibfnamefont {C.~P.}\ \bibnamefont
  {Frahm}},\ }\href {http://aapt.scitation.org/doi/10.1119/1.13127} {\bibfield
  {journal} {\bibinfo  {journal} {American Journal of Physics}\ }\textbf
  {\bibinfo {volume} {51}},\ \bibinfo {pages} {826} (\bibinfo {year}
  {1983})}\BibitemShut {NoStop}%
\bibitem [{\citenamefont {Franklin}(2010)}]{Franklin2010}%
  \BibitemOpen
  \bibfield  {author} {\bibinfo {author} {\bibfnamefont {J.}~\bibnamefont
  {Franklin}},\ }\href {http://arxiv.org/abs/1001.1530
  http://dx.doi.org/10.1119/1.3431987
  http://aapt.scitation.org/doi/10.1119/1.3431987} {\bibfield  {journal}
  {\bibinfo  {journal} {American Journal of Physics}\ }\textbf {\bibinfo
  {volume} {78}},\ \bibinfo {pages} {1225} (\bibinfo {year}
  {2010})}\BibitemShut {NoStop}%
\bibitem [{\citenamefont {Abramowitz}\ and\ \citenamefont
  {Stegun}(1970)}]{Abramowitz1970}%
  \BibitemOpen
  \bibfield  {author} {\bibinfo {author} {\bibfnamefont {M.}~\bibnamefont
  {Abramowitz}}\ and\ \bibinfo {author} {\bibfnamefont {I.~A.}\ \bibnamefont
  {Stegun}},\ }\href@noop {} {\emph {\bibinfo {title} {{Handbook of
  mathematical functions: with formulas, graphs, and mathematical tables}}}}\
  (\bibinfo  {publisher} {Dover Publications},\ \bibinfo {year}
  {1970})\BibitemShut {NoStop}%
\bibitem [{\citenamefont {Sipe}(1987)}]{Sipe1987}%
  \BibitemOpen
  \bibfield  {author} {\bibinfo {author} {\bibfnamefont {J.~E.}\ \bibnamefont
  {Sipe}},\ }\href
  {https://www.osapublishing.org/abstract.cfm?URI=josab-4-4-481} {\bibfield
  {journal} {\bibinfo  {journal} {Journal of the Optical Society of America B}\
  }\textbf {\bibinfo {volume} {4}},\ \bibinfo {pages} {481} (\bibinfo {year}
  {1987})}\BibitemShut {NoStop}%
\bibitem [{\citenamefont {Toma{\v{s}}}(1995)}]{Tomas1995}%
  \BibitemOpen
  \bibfield  {author} {\bibinfo {author} {\bibfnamefont {M.~S.}\ \bibnamefont
  {Toma{\v{s}}}},\ }\href {http://link.aps.org/doi/10.1103/PhysRevA.51.2545}
  {\bibfield  {journal} {\bibinfo  {journal} {Physical Review A}\ }\textbf
  {\bibinfo {volume} {51}},\ \bibinfo {pages} {2545} (\bibinfo {year}
  {1995})}\BibitemShut {NoStop}%
\bibitem [{\citenamefont {Archambault}\ \emph {et~al.}(2009)\citenamefont
  {Archambault}, \citenamefont {Teperik}, \citenamefont {Marquier},\ and\
  \citenamefont {Greffet}}]{Archambault2009}%
  \BibitemOpen
  \bibfield  {author} {\bibinfo {author} {\bibfnamefont {A.}~\bibnamefont
  {Archambault}}, \bibinfo {author} {\bibfnamefont {T.~V.}\ \bibnamefont
  {Teperik}}, \bibinfo {author} {\bibfnamefont {F.}~\bibnamefont {Marquier}}, \
  and\ \bibinfo {author} {\bibfnamefont {J.~J.}\ \bibnamefont {Greffet}},\
  }\href {http://link.aps.org/doi/10.1103/PhysRevB.79.195414} {\bibfield
  {journal} {\bibinfo  {journal} {Physical Review B}\ }\textbf {\bibinfo
  {volume} {79}},\ \bibinfo {pages} {195414} (\bibinfo {year}
  {2009})}\BibitemShut {NoStop}%
\bibitem [{\citenamefont {Henkel}\ \emph {et~al.}(1999)\citenamefont {Henkel},
  \citenamefont {P{\"{o}}tting},\ and\ \citenamefont {Wilkens}}]{Henkel1999}%
  \BibitemOpen
  \bibfield  {author} {\bibinfo {author} {\bibfnamefont {C.}~\bibnamefont
  {Henkel}}, \bibinfo {author} {\bibfnamefont {S.}~\bibnamefont
  {P{\"{o}}tting}}, \ and\ \bibinfo {author} {\bibfnamefont {M.}~\bibnamefont
  {Wilkens}},\ }\href {http://link.springer.com/10.1007/s003400050823}
  {\bibfield  {journal} {\bibinfo  {journal} {Applied Physics B}\ }\textbf
  {\bibinfo {volume} {69}},\ \bibinfo {pages} {379} (\bibinfo {year}
  {1999})}\BibitemShut {NoStop}%
\bibitem [{\citenamefont {Langsjoen}\ \emph {et~al.}(2012)\citenamefont
  {Langsjoen}, \citenamefont {Poudel}, \citenamefont {Vavilov},\ and\
  \citenamefont {Joynt}}]{Langsjoen2012}%
  \BibitemOpen
  \bibfield  {author} {\bibinfo {author} {\bibfnamefont {L.~S.}\ \bibnamefont
  {Langsjoen}}, \bibinfo {author} {\bibfnamefont {A.}~\bibnamefont {Poudel}},
  \bibinfo {author} {\bibfnamefont {M.~G.}\ \bibnamefont {Vavilov}}, \ and\
  \bibinfo {author} {\bibfnamefont {R.}~\bibnamefont {Joynt}},\ }\href
  {http://link.aps.org/doi/10.1103/PhysRevA.86.010301} {\bibfield  {journal}
  {\bibinfo  {journal} {Physical Review A}\ }\textbf {\bibinfo {volume} {86}},\
  \bibinfo {pages} {010301} (\bibinfo {year} {2012})}\BibitemShut {NoStop}%
\bibitem [{\citenamefont {Langsjoen}\ \emph {et~al.}(2014)\citenamefont
  {Langsjoen}, \citenamefont {Poudel}, \citenamefont {Vavilov},\ and\
  \citenamefont {Joynt}}]{Langsjoen2014}%
  \BibitemOpen
  \bibfield  {author} {\bibinfo {author} {\bibfnamefont {L.~S.}\ \bibnamefont
  {Langsjoen}}, \bibinfo {author} {\bibfnamefont {A.}~\bibnamefont {Poudel}},
  \bibinfo {author} {\bibfnamefont {M.~G.}\ \bibnamefont {Vavilov}}, \ and\
  \bibinfo {author} {\bibfnamefont {R.}~\bibnamefont {Joynt}},\ }\href
  {http://link.aps.org/doi/10.1103/PhysRevB.89.115401} {\bibfield  {journal}
  {\bibinfo  {journal} {Physical Review B}\ }\textbf {\bibinfo {volume} {89}},\
  \bibinfo {pages} {115401} (\bibinfo {year} {2014})}\BibitemShut {NoStop}%
\bibitem [{\citenamefont {Lalanne}\ and\ \citenamefont
  {Hugonin}(2006)}]{Lalanne2006}%
  \BibitemOpen
  \bibfield  {author} {\bibinfo {author} {\bibfnamefont {P.}~\bibnamefont
  {Lalanne}}\ and\ \bibinfo {author} {\bibfnamefont {J.~P.}\ \bibnamefont
  {Hugonin}},\ }\href {http://www.nature.com/doifinder/10.1038/nphys364}
  {\bibfield  {journal} {\bibinfo  {journal} {Nature Physics}\ }\textbf
  {\bibinfo {volume} {2}},\ \bibinfo {pages} {551} (\bibinfo {year}
  {2006})}\BibitemShut {NoStop}%
\bibitem [{\citenamefont {Liu}\ and\ \citenamefont {Lalanne}(2008)}]{Liu2008}%
  \BibitemOpen
  \bibfield  {author} {\bibinfo {author} {\bibfnamefont {H.}~\bibnamefont
  {Liu}}\ and\ \bibinfo {author} {\bibfnamefont {P.}~\bibnamefont {Lalanne}},\
  }\href {http://www.nature.com/doifinder/10.1038/nature06762} {\bibfield
  {journal} {\bibinfo  {journal} {Nature}\ }\textbf {\bibinfo {volume} {452}},\
  \bibinfo {pages} {728} (\bibinfo {year} {2008})}\BibitemShut {NoStop}%
\bibitem [{\citenamefont {Lalanne}(2009)}]{Lalanne2009}%
  \BibitemOpen
  \bibfield  {author} {\bibinfo {author} {\bibfnamefont {P.}~\bibnamefont
  {Lalanne}},\ }\href
  {http://linkinghub.elsevier.com/retrieve/pii/S0167572909000478} {\bibfield
  {journal} {\bibinfo  {journal} {Surface Science Reports}\ }\textbf {\bibinfo
  {volume} {64}},\ \bibinfo {pages} {453} (\bibinfo {year} {2009})}\BibitemShut
  {NoStop}%
\end{thebibliography}


\end{document}